\documentclass[aps,pra,reprint,longbibliography,superscriptaddress]{revtex4-1}
\usepackage{mathtools}
\usepackage{bbold}
\usepackage{physics}
\usepackage{amsmath}
\usepackage{xcolor}

\begin{document}

\title{Effect of coordination on topological phases on self-similar structures}

\author{Saswat Sarangi}
\affiliation{Max-Planck-Institut f\"{u}r Physik komplexer Systeme,
D-01187 Dresden, Germany}

\author{Anne E. B. Nielsen}
\affiliation{Max-Planck-Institut f\"{u}r Physik komplexer Systeme,
D-01187 Dresden, Germany}
\affiliation{Department of Physics and Astronomy,
Aarhus University, DK-8000 Aarhus C, Denmark}

\begin{abstract}
Topologically non-trivial phases have recently been reported on self-similar structures. Here, we investigate the effect of local structure, specifically the role of the coordination number, on the topological phases on self-similar structures embedded in two dimensions. We study a geometry dependent model on two self-similar structures having different coordination numbers, constructed from the Sierpinski Gasket. For different non-spatial symmetries present in the system, we numerically study and compare the phases on both the structures. We characterize these phases by the localization properties of the single-particle states, their robustness to disorder, and by using a real-space topological index. We find that both the structures host topologically non-trivial phases and the phase diagrams are different on the two structures. This suggests that, in order to extend the present classification scheme of topological phases to non-periodic structures, one should use a framework which explicitly takes the coordination of sites into account.
\end{abstract}

\maketitle

\section{Introduction}
After the discovery of the quantum Hall effect,  the study of topological phases has been one of the leading research areas in condensed matter physics. In non-interacting electronic systems, topologically non-trivial phases are usually identified by the presence of gapless boundary modes highly robust to weak disorders, and are characterized by relevant topological invariants \cite{Kane2005,Fu2007}. These phases are well understood for translationally invariant systems, as the presence of a well-defined momentum eigenbasis gives a natural setting to describe the topology of bulk wavefunctions. Systematic classification of topological phases on non-interacting translationally invariant systems has been done in terms of both non-spatial and spatial symmetries \cite{Schnyder2008,Kitaev2009,Fu2011,Ludwig2015,Slager2013,Chiu2016}.
\newline

Although translational invariance is a necessary condition for the presence of a well-defined momentum eigenbasis, it turns out that this is not a necessary condition for the existence of topological phases. Topological phases have been reported in quasiperiodic, quasicrystalline, and amorphous systems  \cite{Duncan2020,Agarwala2017,Mitchell2018} which only preserve the notion of a well-defined ``bulk" and ``boundary", as defined in regular lattice systems with open ``boundary". Also recently, properties associated with topological phases have been reported on finite truncations of fractals like the Sierpinski Gasket and Sierpinski Carpet \cite{Agarwala2018,Brzezinska2018,Fremling2019,Pai2019,Iliasov2020} which even lack this notion of ``bulk" and ``boundary". Although there have been some speculations \cite{Agarwala2018, Brzezinska2018, Pai2019}, the factors affecting the topological properties of systems without a precise bulk-boundary distinction, are yet to be clearly identified. In an attempt to identify one such factor, here we study the effect of coordination on the topological properties of non-interacting Hamiltonians on self-similar structures. 
\newline

The way the sites are coordinated locally on a lattice plays an important role in determining which topological phases the lattice can host. To see this, consider a general two-orbital nearest-neighbor tight-binding model on a 2D Bravais lattice, similar to what is considered in \cite{Fu2011}, given by 
\begin{equation}
H_{tb}=\sum_{\textbf{R},<\textbf{r}>,\alpha,\beta}t(\textbf{r})(\psi^{\dagger}_{\alpha}(\textbf{R})f(\theta_{\textbf{r}})\psi_{\beta}(\textbf{R}+\textbf{r})),
\end{equation}
where \textbf{R} specifies the position vectors for the sites, \textbf{r} specifies the relative vectors between two sites, $\lbrace\alpha$, $\beta\rbrace$ label the two orbitals, and $(\cos(\theta_{\textbf{r}}), \sin(\theta_{\textbf{r}}))=\textbf{r}/|\textbf{r}|$. The function $f(\theta_{\textbf{r}})$ is any function such that $H_{tb}$ is Hermitian. The matrix elements of the corresponding Bloch Hamiltonian $H_{tb}(\textbf{k})$, which essentially determine the band topology, encode the information about the local structure of the lattice as they involve a sum over all nearest neighbors. This is how local properties like coordination comes into the picture. As the form of $H_{tb}$ is entirely determined by the crystal symmetry of the underlying lattice \cite{Fu2011}, crystal symmetries are used for topological classification of such systems. Also, crystal symmetries are known to put constraints on bulk topological invariants \cite{Fang2012}. On some two dimensional lattices, the graph of the model, formed by identifying the sites as the vertices and the non-zero hoppings as the edges, forms a regular tiling of the two dimensional space. For such cases, the coordination number is uniquely determined by the crystal symmetry and the coordination number is hence not a separate variable that could influence the topological properties. Examples of such cases are nearest neighbor models on triangular, square and hexagonal lattices.  But on self-similar structures, to the best of our knowledge, no such correspondence has been established between coordination and spatial symmetries. It is hence an open question whether a change only in the local coordination of the sites can affect the topological phases on self-similar structures.
\newline

The idea of coordination is also crucial for the distinction between ``bulk'' and ``boundary'' on regular lattices. But, self-similar structures lack a clear distinction between bulk and boundary. However coordination number, and hence the notion of coordination, is well defined for self-similar structures, as those are special graphs like regular lattices. For this study, we first construct two different self-similar structures from the Sierpinski Gasket (SG), with different coordination numbers, which have the same Hausdorff dimension. We then numerically study a geometry dependent non-interacting nearest neighbor Hamiltonian on both structures by looking at certain observables of interest. 
\newline

The rest of the paper is organized as follows. In  section \ref{Structures}, we describe the construction of the two different fractal structures mentioned in the previous paragraph. We define the model Hamiltonian and the observables we are looking at in section \ref{Model}.  In section \ref{Results}, we present and compare various properties of the Hamiltonian on both the structures. Finally, in section \ref{Conclusion}, we conclude with a summary of our results and discuss some of the remaining open questions on the subject.

\section{Construction of Fractal Structures}\label{Structures}
\begin{figure}
\includegraphics[scale=0.15]{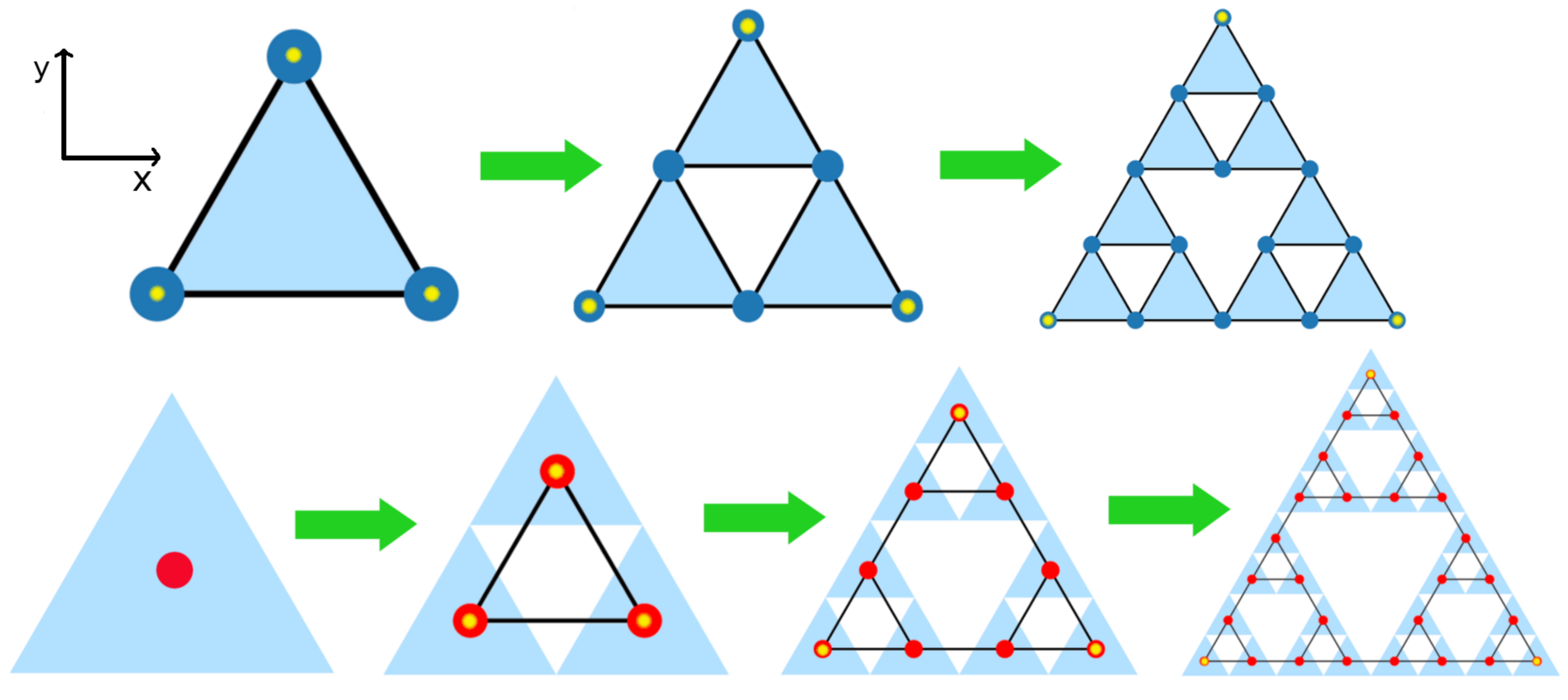}
\caption{\footnotesize{Schematics for the construction of the self-similar structures. The shaded regions are the finite truncations of the SG for different generations. The blue and red dots indicate the positions of the sites for the structures SG-4 and SG-3, respectively. The black solid lines represent the bonds between the sites. The three corner sites are marked with an additional yellow dot in both the structures.}}
\label{Construction_SG}
\end{figure}

It is possible to construct various graphs on the SG, but for simplicity, we chose to focus on self-similar graphs which are equi-coordinated, except at the corner sites. One can construct self-similar structures with coordination numbers 3 and 4 as illustrated in Fig.\ \ref{Construction_SG}, and of course also with coordination number zero which is trivial, but we have not found equi-coordinated graphs with other coordination numbers. We hence focus on the structures in Fig.\ \ref{Construction_SG} in the following. 
\newline

 First we construct the SG by a recursive procedure starting from an equilateral triangle. We divide it into four equilateral triangles of equal area, remove the central triangle, and repeat the procedure infinitely for each of the remaining triangles. We call the structure generated after $g$ iterations for `SG with generation $g$' and the triangles removed in a particular iteration for `triangles belonging to generation $g$'.
\newline
 
For the first structure (shown in Fig.\ \ref{Construction_SG}), we identify the vertices of the triangles in each generation of the SG with the sites, and the edges with the bonds. This gives a self-similar structure in which, all sites except the three corner sites (marked in yellow in Fig.\ \ref{Construction_SG}), have coordination number 4. We denote this structure by `SG-4'. Tight binding  models on this type of structure have been extensively studied using real space renormalization methods \cite{Domany1983,Kimball1998,Rammal1982}.
\newline

 For the second structure (also shown in Fig.\ \ref{Construction_SG}), we identify the centroids of the smallest triangles in each generation of the SG with the sites, and connect the nearest neighbors. This also gives a self-simliar structure. But in this case, in each generation, all sites except the three corner sites, have coordination number 3.  We denote this structure by `SG-3'. Notice that the first generation of the SG-4 is obtained from the zeroth generation of the SG, whereas the first generation of the SG-3 is obtained from the first generation of the SG. 
\newline

\begin{figure}
\includegraphics[scale=0.18]{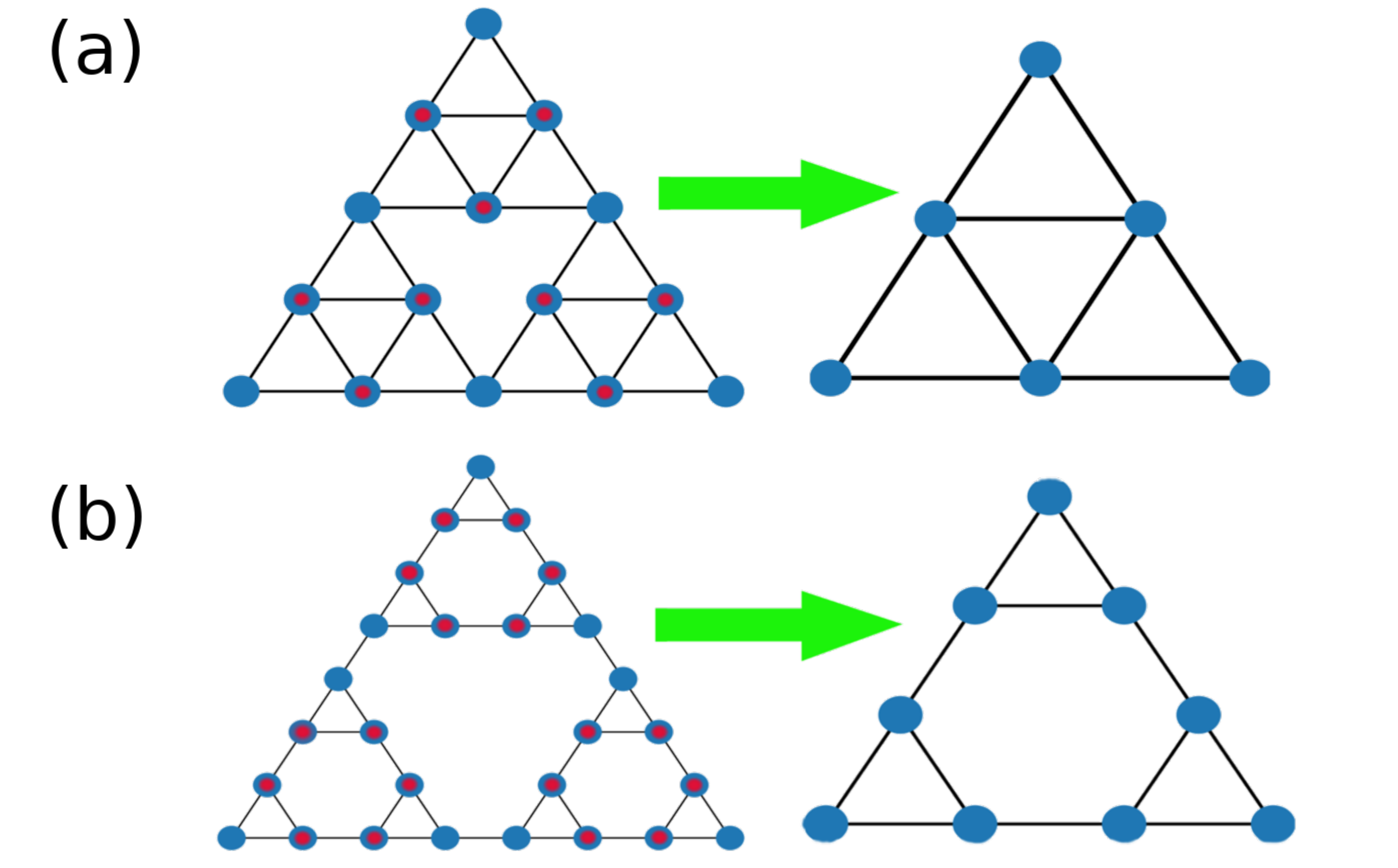}
\caption{\footnotesize{Schematics showing the self-similarity of (a) SG-4 and (b) SG-3. The sites of the 3rd generation are marked in red for both the structures. When these sites are removed, the remaining structure resembles that of the 2nd generation.}}
\label{Decimation_SG}
\end{figure}

Due to the self-similar nature of the SG-3 and the SG-4, for each structure, we can remove certain specific sites from a given generation `$g$' so that the structure with the remaining sites resembles that of generation `$g-1$'. For each structure, we term these specific sites as the `sites of generation $g$'. This is illustrated in Fig.\ \ref{Decimation_SG}. In both the structures, in each generation, only the three corner sites of the SG-3 and the SG-4 are two coordinated, but we expect this to not affect the physics when we are far from the corner sites. Notice that both the structures have the same Hausdorff dimension as the SG. For numerical calculations, we carry out the constructions mentioned above, but with a finite number of iterations for the SG, which gives us structures with finite number of sites for the SG-3 and the SG-4.

\section{Model and Approach}\label{Model}
We study the fermionic, generalised Bernevig-Hughes-Zhang (\textit{BHZ}) model, on the self-similar structures mentioned in  section \ref{Structures}. We choose to study this model because the \textit{BHZ} model is known to host topologically non-trivial quantum spin Hall insulating phases on translationally invariant lattice systems \cite{Bernevig2006}. Also, this model can be easily generalized to make it depend on the geometry of the underlying motif \cite{Agarwala2017, Agarwala2018}. We define the model in the following way. Each site has two orbital degrees of freedom, denoted by $\alpha=\lbrace c,d\rbrace$, and two spin degrees of freedom, denoted by $\sigma=\pm 1$. 
We consider only nearest neighbor hopping. The Hamiltonian is given by
\begin{multline}\label{4_band_ham}
\hat{\mathbb{H}}_{BHZ}=M\sum_{j \sigma}{\hat{\psi}}^{\dagger}_{j \sigma}\tau_{z}{\hat{\psi}}_{j \sigma}-t\sum_{<jk>,\sigma}{\hat{\psi}}^{\dagger}_{j\sigma}\tau_{z}{\hat{\psi}}_{k \sigma} \\ - \lambda \sum_{<jk>,\sigma}{\hat{\psi}}^{\dagger}_{j\sigma} {\sigma }{\tilde{\mathcal{T}}}_{jk,\sigma}{\hat{\psi}}_{k \sigma}
\end{multline}
where ${\hat{\psi}}^{\dagger}_{j\sigma}=({\hat{c}}^{\dagger}_{j\sigma},{\hat{d}}^{\dagger}_{j\sigma})$,  $\lbrace \tau_x,\tau_y,\tau_z \rbrace$ are the Pauli matrices for the orbital degrees of freedom, and $\tilde{\mathcal{T}}_{jk,\sigma}$ is given by
\[
\tilde{\mathcal{T}}_{jk,\sigma}=
\begin{pmatrix}
0 & ie^{-i\sigma \theta_{jk}} \\ie^{i\sigma \theta_{jk}}  & 0
\end{pmatrix}.
\]
Here $\theta_{jk}$ denotes the angle made by the vector from the $j$th site to the $k$th site, with the $x$ axis. $M$ denotes the on-site energy. The real, non-negative numbers $t$ and $\lambda$ denote the hopping strengths for hopping between the same orbitals  and different orbitals of nearest neighbor sites, respectively. 
\newline

The two $\sigma$ sectors are decoupled from each other and are time reversal partners of each other, so it suffices to study the model for one value of $\sigma$. Here we look only at the $\sigma=1$ sector and hence the respective two-orbital Hamiltonian is given by
\begin{equation}\label{2_band_ham}
\hat{\mathbb{H}}=M\sum_{j }{\hat{\psi}}^{\dagger}_{j}\tau_{z}{\hat{\psi}}_{j}-t\sum_{<jk>}{\hat{\psi}}^{\dagger}_{j}\tau_{z}{\hat{\psi}}_{k} -\lambda \sum_{<jk>}{\hat{\psi}}^{\dagger}_{j} {\mathcal{T}}_{jk}{\hat{\psi}}_{k},
\end{equation}
\[
{\mathcal{T}}_{jk}=
\begin{pmatrix}
0 & ie^{-i \theta_{jk}} \\ie^{i \theta_{jk}}  & 0
\end{pmatrix}.
\]
The model in Eq.\ \eqref{2_band_ham} is the generalized \textit{half-BHZ} model and is known to host topological phases on square and triangular lattices. For $\lambda=t$, this model hosts two distinct topological phases on a square lattice with Chern number 1 and -1 \cite{Bernevig2006, Asboth2016}. However, on a triangular lattice, this model hosts a different topological phase with Chern number -2, along with a trivial phase and a topological phase with Chern number 1 \cite{Agarwala2018}. This is a classic example where different coordination numbers in different lattices result in emergence of different topological phases. Also, for $t=\lambda=1/2$, this model has been studied on a fractal structure  which is closely related to SG-4, but with different boundary conditions \cite{Agarwala2018}.
\newline

We numerically study the systems by primarily looking at the localization, dynamics and the topological nature of the single-particle states at half-filling. For the numerical computations, we use KWANT code \cite{Groth2014}. A single particle state denoted by label $n$ can be written as
\begin{equation}
\ket{\psi_n}=\sum_{j\alpha} \varphi_{n,j\alpha}\ket{j\alpha}
\end{equation}
where $\left\lbrace\ket{j}\right\rbrace$ denotes the basis vectors in the site basis. We study the localization of single particle states by looking at the density at any site $j$, given by
\begin{equation}
 \rho_n(j)=\sum_{\alpha}\abs{\varphi_{n,j\alpha}}^2.
\end{equation} 
Given that it is unclear how to have a sharp distinction between bulk and edge states in the case of fractal systems, we  define `bulk-like' and `edge-like' states as follows. An eigenstate is a bulk-like state if it has finite probability density on sites which enclose the triangles belonging to more or less every generation of the SG. On the other hand, an eigenstate is an edge-like state, if it is localized on sites which enclose the triangles belonging entirely to a particular generation of the SG.
\newline

\begin{figure}
\raisebox{22mm}{\textbf{(a)}}
\includegraphics[scale=0.2]{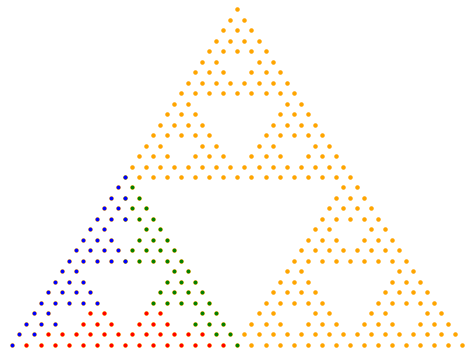}
\hfill
\raisebox{22mm}{\textbf{(b)}}
\includegraphics[scale=0.2]{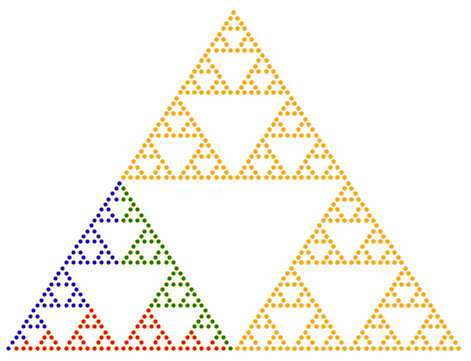}
  \caption{\footnotesize{Partitions of the $6$th generation of \textbf{(a)} SG-4 and \textbf{(b)} SG-3, for the real space Chern number calculation. The regions \textbf{A},\textbf{B} and \textbf{C} are marked in red, green, and blue respectively. The subsection is \textbf{X}$=$\textbf{A}$\cup $\textbf{B}$\cup $\textbf{C}}}\label{SS}
\end{figure}

We use Kitaev's topological index to study the topological properties of the systems, which relies solely on the real space description of the system \citep{Kitaev2006}. This has been used in the literature to study the topological phases on self-similar structures \cite{Brzezinska2018, Fremling2019}. We first choose a subsection \textbf{X} of the fractal and divide it into three parts, \textbf{A}, \textbf{B} and \textbf{C}, as shown in Fig. \ref{SS}. We use the following expression for the real space Chern number
\begin{equation}\label{LCN}
\nu (P)=12 \pi i (\textrm{Tr}(APBPCP)-\textrm{Tr}(APCPBP))
\end{equation}
where $P=\sum_{k}\ket{\psi_{k}}\bra{\psi_{k}}$ is the projector onto the desired eigen states. $A,B,C$ are diagonal matrices with 
\begin{equation}
A=\tilde{A}\otimes\mathbb{1}_{N_{orb}}~~~
B=\tilde{B}\otimes\mathbb{1}_{N_{orb}}~~~
C=\tilde{C}\otimes\mathbb{1}_{N_{orb}}
\end{equation}
where $\tilde{A},\tilde{B}, \tilde{C}$ denote the projectors into the sectors \textbf{A},\textbf{B},\textbf{C} (as shown in Fig.\ \ref{SS}) respectively, and $N_{orb}$ is the number of orbitals per site which is 2 in this case. 
\newline

We also check the dynamics of the states close to the Fermi energy. To do this, we project a single particle state, initially localized in the $c$ orbital of one of the sites of the fractal, onto a part of the eigenbasis  defined by $E_{min}<E<E_{F}$,  and then time evolve under $\hat{\mathbb{H}}$. Here, $E$ denotes the eigen-energies of the Hamiltonian and $E_{F}$ denotes the Fermi energy. $E_{min}$ is chosen such that the energy range, $(E_{min},E_{F})$, is small enough to look at the states near the Fermi energy but also large enough to encompass all the edge-like states below the Fermi energy. All the computations have been done with $E_{min}=-0.5$. Apart from this, we check whether the dynamics change in the presence of disorder. To do this we add an extra onsite Anderson disorder term to the Hamiltonian of the form
\begin{equation}
\hat{\mathbb{W}}=\sum_{j}{\hat{\psi}}^{\dagger}_{j}\tilde{W}_{j}{\hat{\psi}}_{j}
\end{equation}
where $\tilde{W}_{j}=\textrm{diag}(\epsilon^{c}_{j}, \epsilon^{d}_{j})$ and $\epsilon^{c}_{j}, \epsilon^{d}_{j}$ are random numbers drawn from a uniform random distribution with mean $\mu=0$ and variance $W$. The total Hamiltonian under which the system is time evolved then becomes $\hat{\mathbb{H}}_{dis}=\hat{\mathbb{H}}+\hat{\mathbb{W}}$.

\section{Results}\label{Results}
The Hamiltonian in Eq.\ \eqref{2_band_ham}, can be rewritten in the following block form in $orbital \otimes site$ notation
\begin{equation}\label{orb_ham}
\hat{\mathbb{H}}=\hat{\Psi}^{\dagger}\mathbb{H} \hat{\Psi},
\end{equation}
\[\hat{\Psi}=
\begin{pmatrix}
\hat{C} \\ \hat{D}
\end{pmatrix}, ~~~~
\mathbb{H}=
\begin{pmatrix}
M-tH & \lambda\Delta \\ \lambda\Delta^{\dagger} & -(M-tH) 
\end{pmatrix}.
\]
Here, $\hat{C}$=${(\hat{c}_1,\hat{c}_2,...,\hat{c}_{N_{s}})}^{\text{T}}$ and $\hat{D}$=${(\hat{d}_1,\hat{d}_2,...,\hat{d}_{N_{s}})}^{\text{T}}$, where $N_{s}$ is the total number of sites. $\Delta_{jk}=-ie^{-i \theta_{jk}}$ and ${H}_{jk}=1$, if $j,k$ are nearest neighbors connected by a bond as shown in Fig.\ \ref{Construction_SG}, and otherwise zero. From Eq.\ \eqref{orb_ham}, it is easy to see that this model has a charge-conjugation symmetry for all values of $M,t,$ and $\lambda$, given by
\begin{equation}\label{C_symmetry}
 P^{-1} \mathbb{H} P= -\mathbb{H}.
\end{equation} 
Here $P=\tau_x \mathcal{K}$, where $\mathcal{K}$ is the complex conjugation operator. A consequence of this symmetry is the spectra being symmetric around zero energy. Apart from this, the Hamiltonian has other non-spatial symmetries for certain specific parameter values. So we break our results into three parts, specifically focusing on three particular parameter regimes, each having different symmetry properties.

\subsection{$t\neq\lambda=0$}
\begin{figure}
\hspace{0cm}
\includegraphics[scale=0.18]{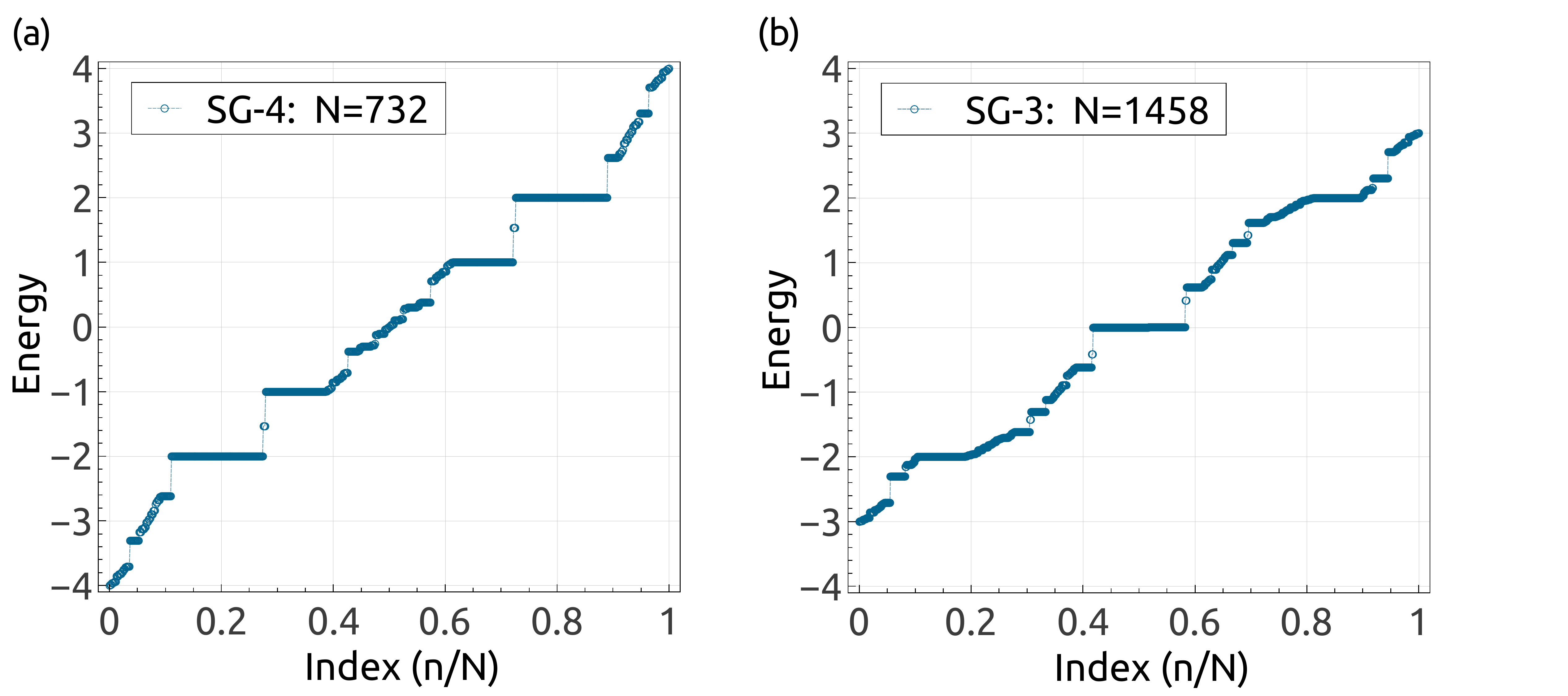}
\caption{\footnotesize{Spectrum of $\mathbb{H}$ for $\lambda=0$, $M=0$, and $t=1$ on (a) SG-4 with $g=6$ and (b) SG-3 with $g=6$. $\text{N}=2N_{s}$ denotes the total number of eigenstates where $N_{s}$ is the total number of sites.}}
\label{Spectra_tb}
\end{figure}

For $\lambda=0$, the Hamiltonian in Eq.\ \eqref{orb_ham} becomes

\begin{equation}\label{Hamiltonian_tb}
\mathbb{H}= \tau_z\otimes (M-tH)
\end{equation}
 which is block diagonal and decouples into two single orbital tight-binding models. This is well studied in the literature on the  SG-4 \cite{Domany1983,Kimball1998,Rammal1982}.  The spectrum of the model (shown in Fig.\ \ref{Spectra_tb}) is symmetric about $E=0$, as expected, due to the charge-conjugation symmetry\ \eqref{C_symmetry} of the model.  It is already known for SG-4 that the spectrum is self similar and has infinitely many gaps in the infinite $g$ limit. We find that the spectrum of SG-3 is also self-similar with infinitely many gaps in the infinite $g$ limit. We confirm this numerically by computing the spectrum for different $g$ values, and analytically by following the renormalisation procedure done in \cite{Domany1983,Kimball1998}. For $M=0$, as seen in Fig.\ \ref{Spectra_tb}(b), we see a very high degeneracy at zero energy in case of SG-3, which is not seen in case of SG-4. The model has the symmetry that $\tau_z$ commutes with the Hamiltonian\ \eqref{Hamiltonian_tb}, but this only gives rise to a twofold degeneracy. The large degeneracy is hence a consequence of the spatial arrangement of the sites in the underlying structure and not due to any non-spatial symmetry of the Hamiltonian. In this particular regime, however, the Hamiltonian does not host any topological phases on either of the structures as $H$ does not host any topological phase. A nonzero mass term $M$, simply opens up a trivial gap in the spectra.

\subsection{$\lambda\neq t=0$}\label{Tsym}
Now we consider the case when we only have the on-site term,  $c\rightarrow d$ hoppings, and $d\rightarrow c$ hoppings. Then the Hamiltonian matrix $\mathbb{H}$ in Eq.\ \eqref{orb_ham} reduces to
\begin{equation}
\mathbb{H}= M\tau_z + \lambda
\begin{pmatrix}
0 & \Delta \\ \Delta^{\dagger} & 0 
\end{pmatrix} ~\overset{def}{=} M\tau_z + \lambda H_{xy}.
\end{equation}
We start by studying $\mathbb{H}$ for $M=0$. We see that every energy level is at least doubly degenerate on both the structures. This is because $H_{xy}$ has an additional orbital symmetry given by $\tau_z H_{xy} \tau_z= -H_{xy}$ along with the charge-conjugation symmetry\ \eqref{C_symmetry}. Hence, the system possesses time-reversal symmetry given by $T^{-1} H_{xy} T= H_{xy}$, where $T=i\tau_y \mathcal{K}$, which results in the Kramers degeneracy. If $\ket\psi={(C,D)}^{\text{T}}$, where $C=(c_1,c_2,..,c_{N_s})^{\text{T}}$ and $D=(d_1,d_2,..,d_{N_s})^{\text{T}}$, is an eigenstate of $H_{xy}$, then $T\ket\psi={(-D^{*},C^{*})}^{\text{T}}$ is also an eigenstate of $H_{xy}$. Also, $\psi$ and $T\psi$ are orthogonal to each other as $\bra\psi\ket{T\psi}=0$.
\newline

\begin{figure}
\hspace{0cm}
\includegraphics[scale=0.175]{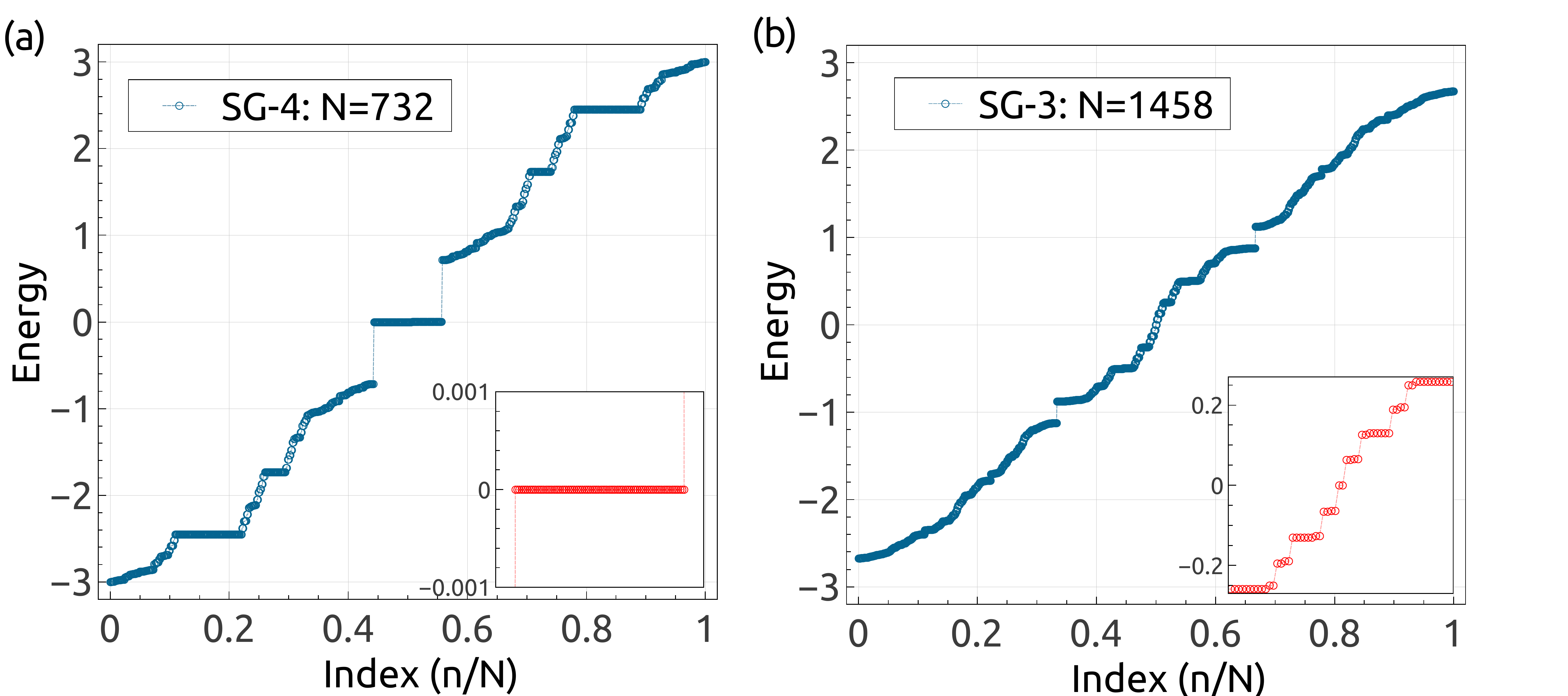}
\caption{\footnotesize{Spectrum of $H_{xy}$ on (a) SG-4 with $g=6$ and (b) SG-3 with $g=6$. $\text{N}=2N_{s}$ denotes the total number of eigenstates where $N_{s}$ is the total number of sites. For (a), the inset shows the highly degenerate levels (flat band) at $E=0$. For (b), the inset shows the whole range of edge-like states near zero energy.}}
\label{Spectra_H_xy}
\end{figure}

\begin{figure}
\includegraphics[scale=0.29]{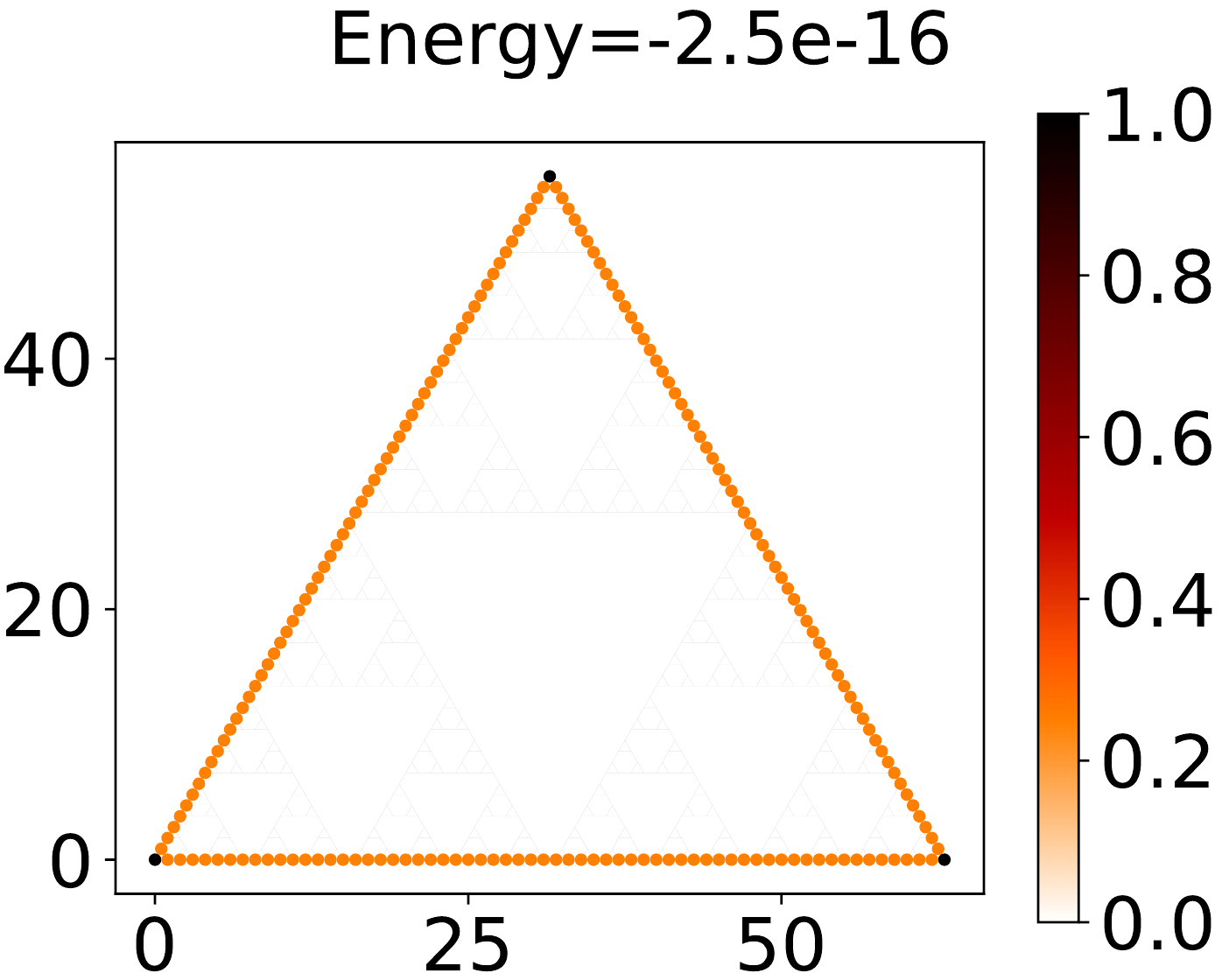}
\includegraphics[scale=0.29]{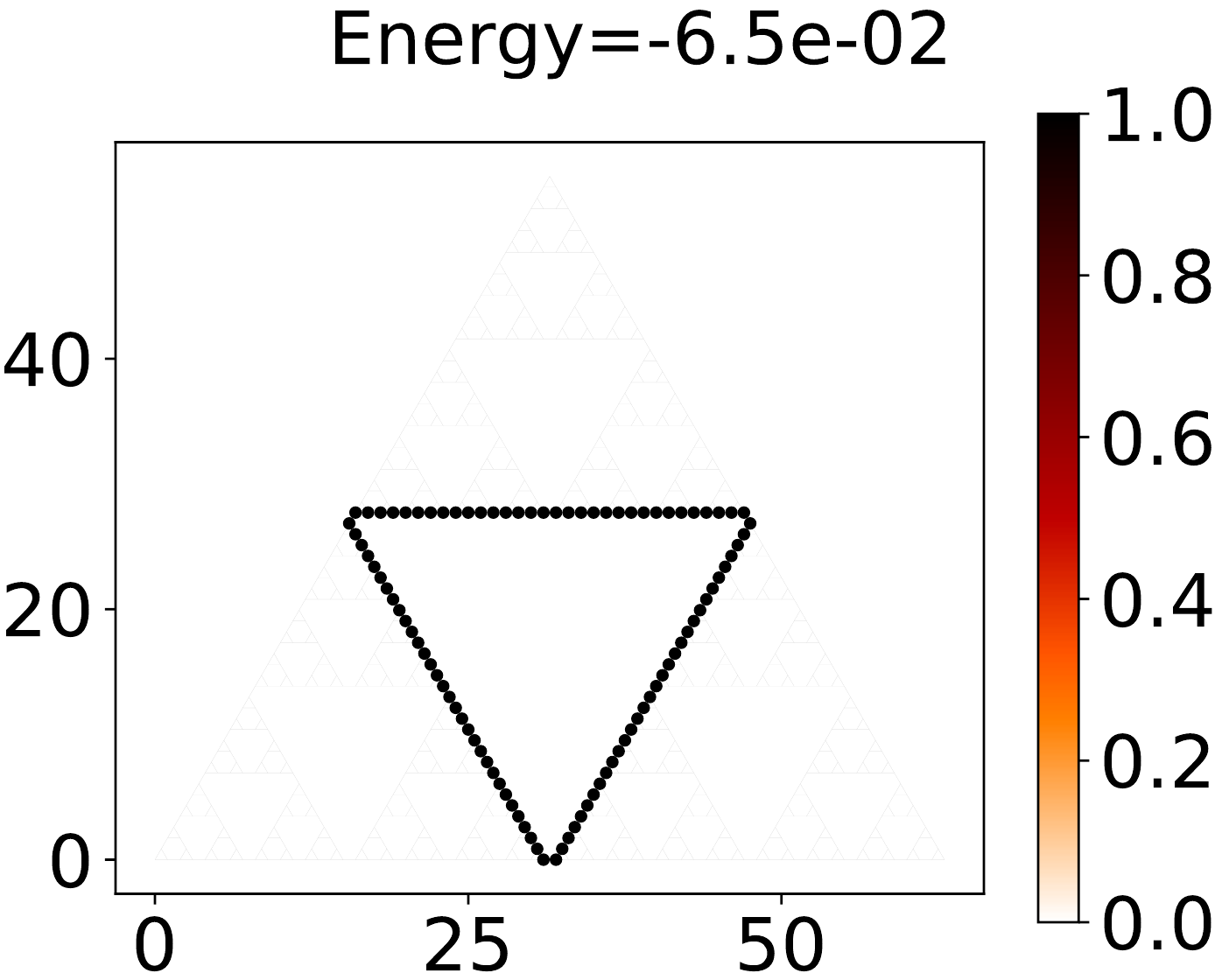}  
\includegraphics[scale=0.29]{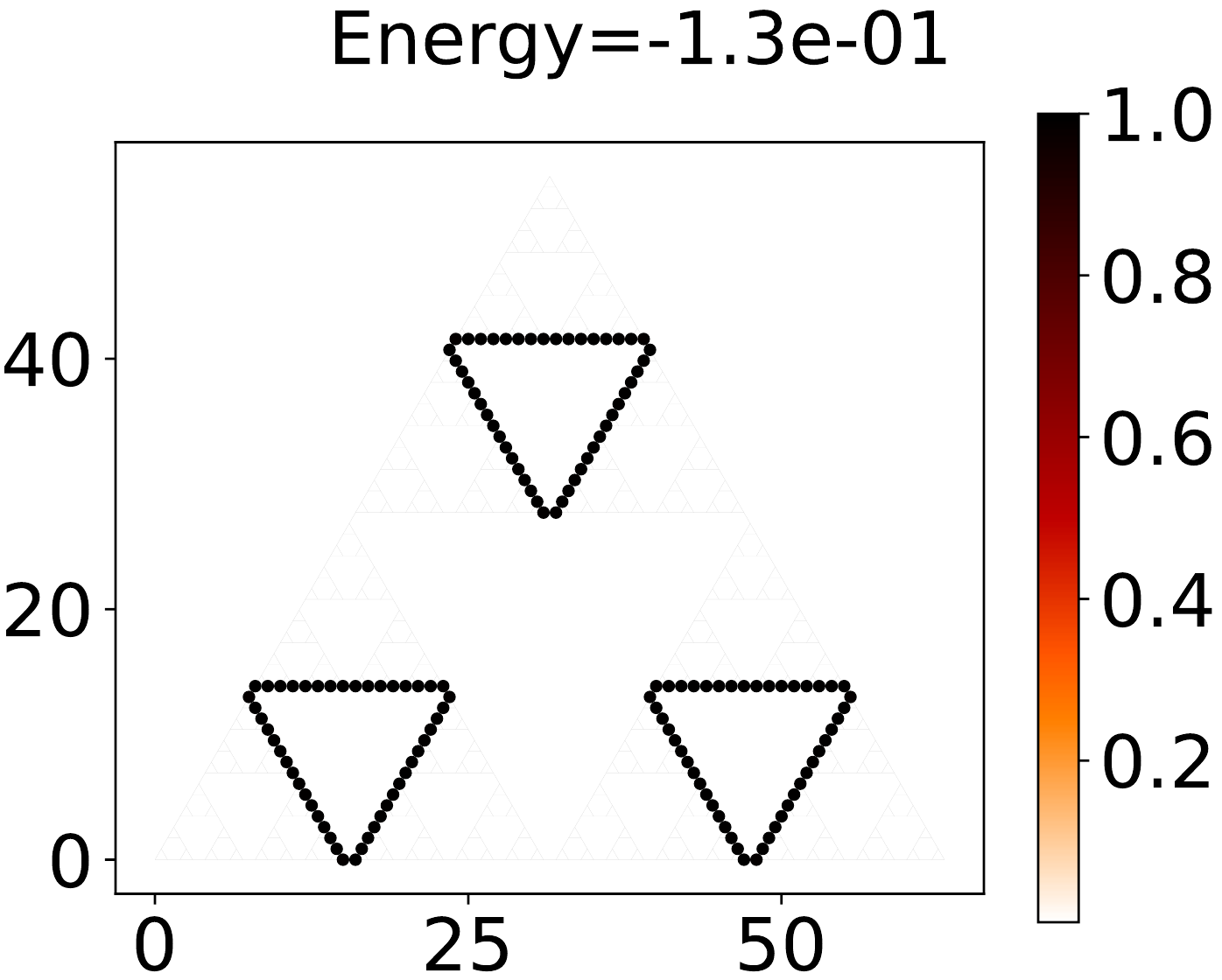}  
\includegraphics[scale=0.29]{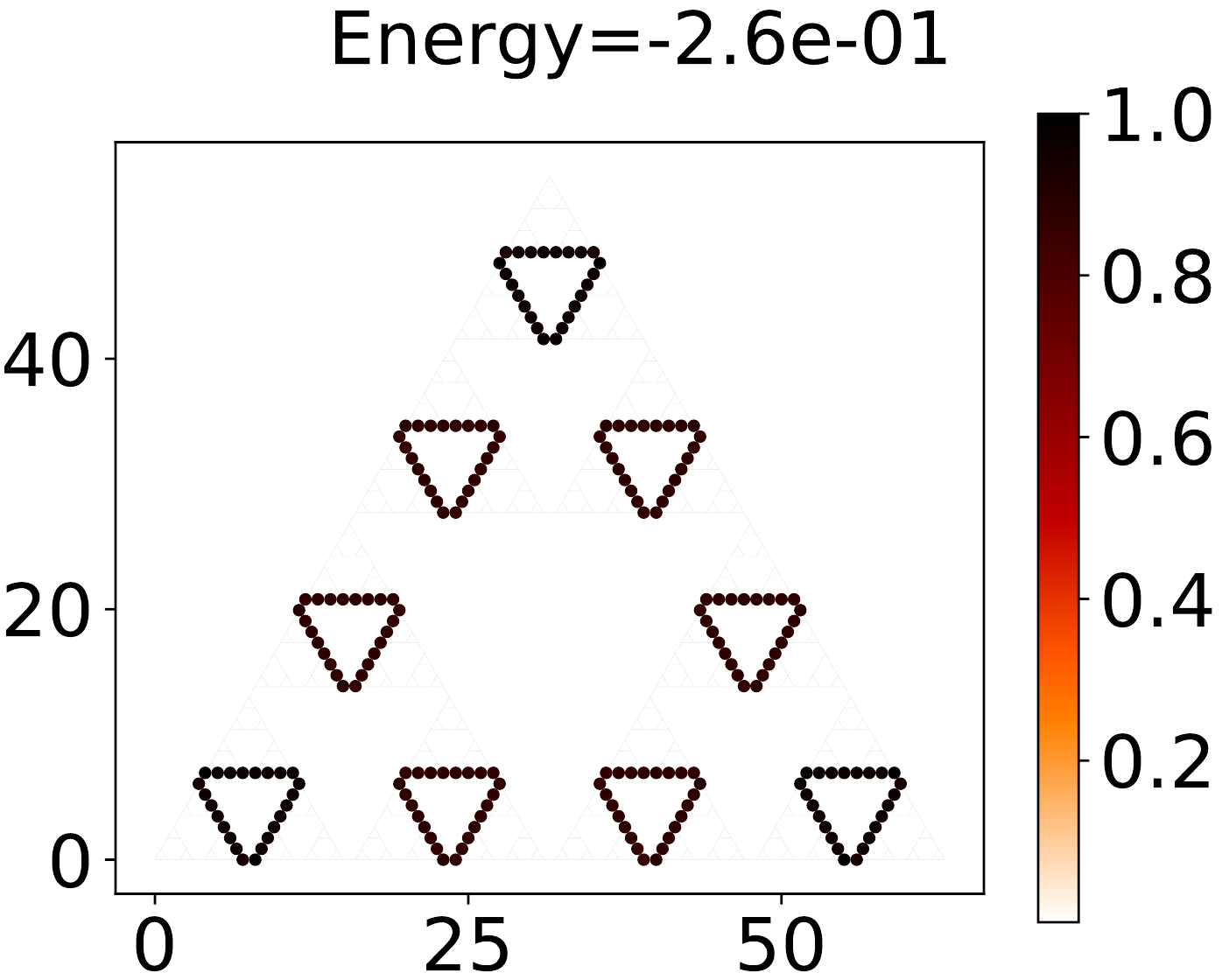}  
\caption{\footnotesize{Few examples of edge-like eigenstates of $H_{xy}$ on SG-3 with $g=6$, close to zero energy. The color bar represents the relative density per site of an eigenstate, $\ket{\psi_n}$, defined by $\rho_n(j) / {\max}(\rho_n(j))$.}}\label{Wv_SG_3_t_0_M_0}
\end{figure}

\begin{figure}
\includegraphics[scale=0.25]{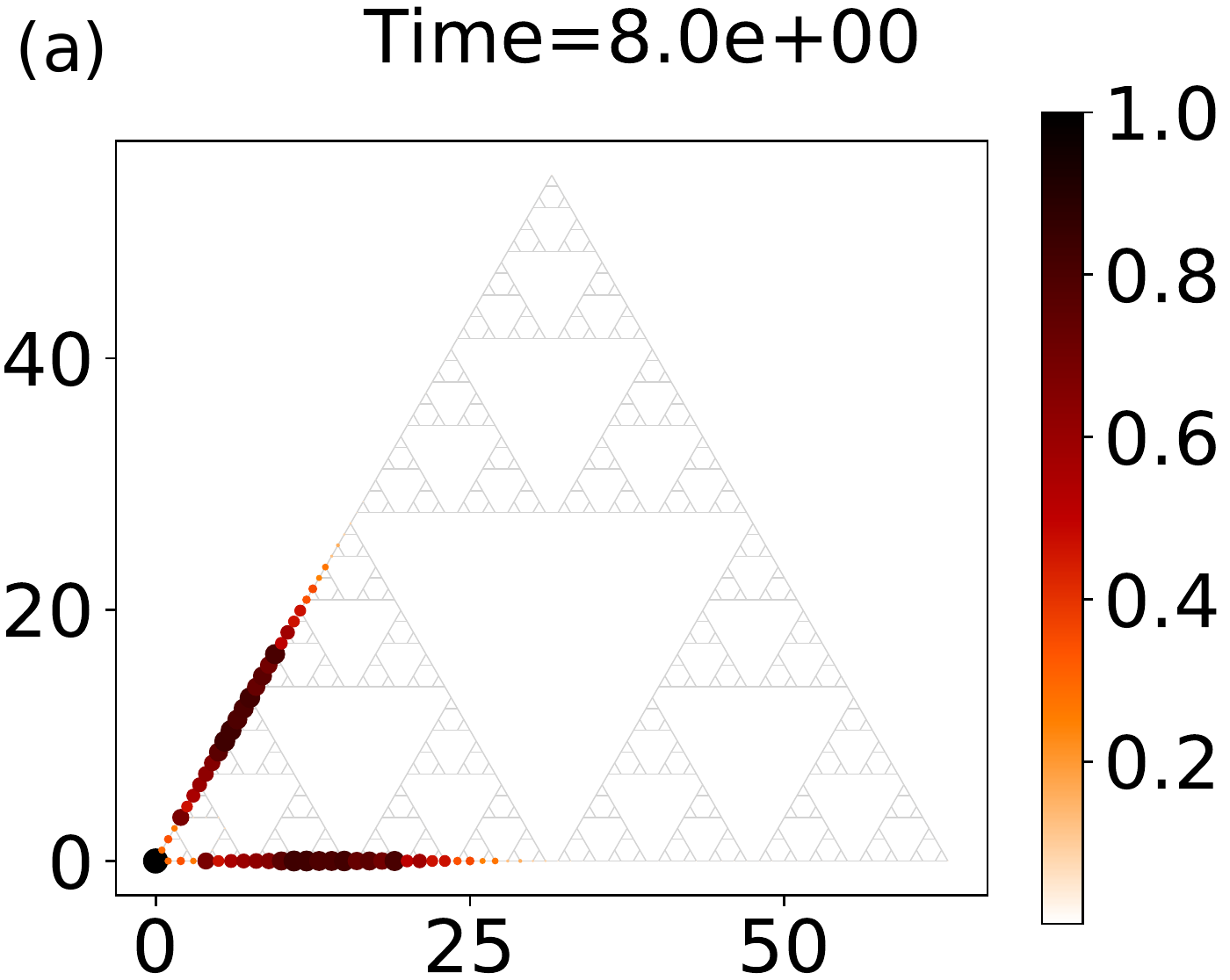}
\includegraphics[scale=0.25]{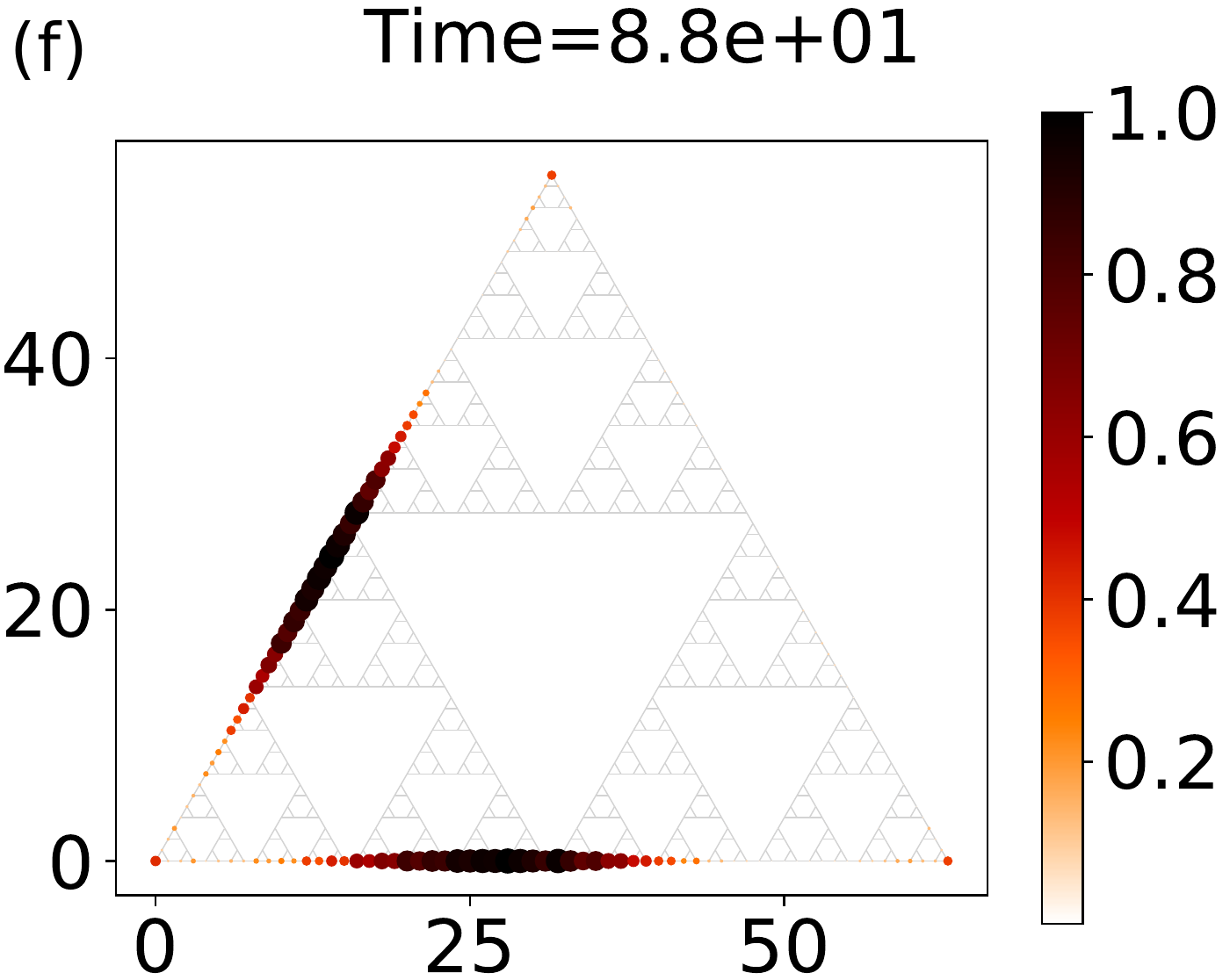}  
\includegraphics[scale=0.25]{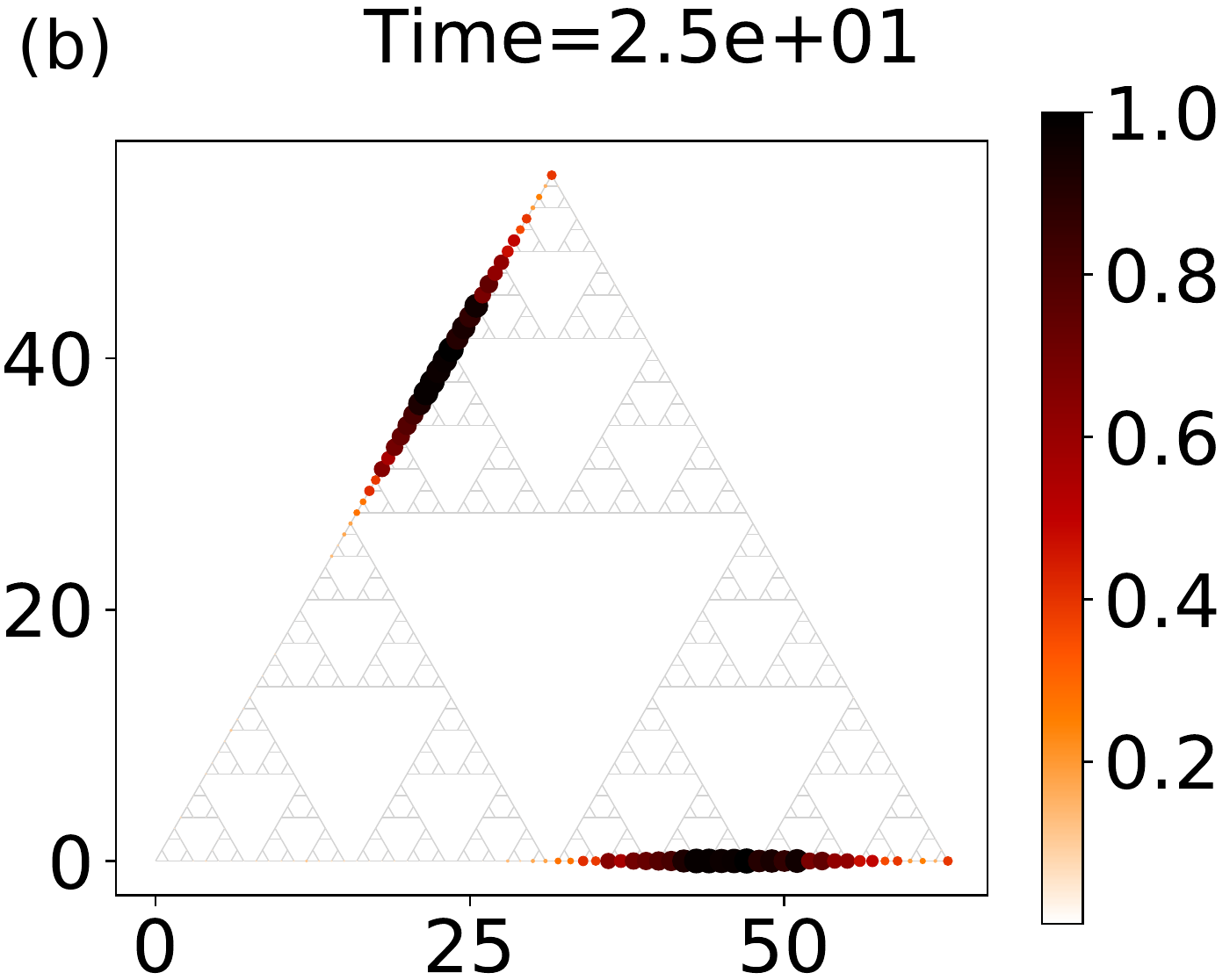}  
\includegraphics[scale=0.25]{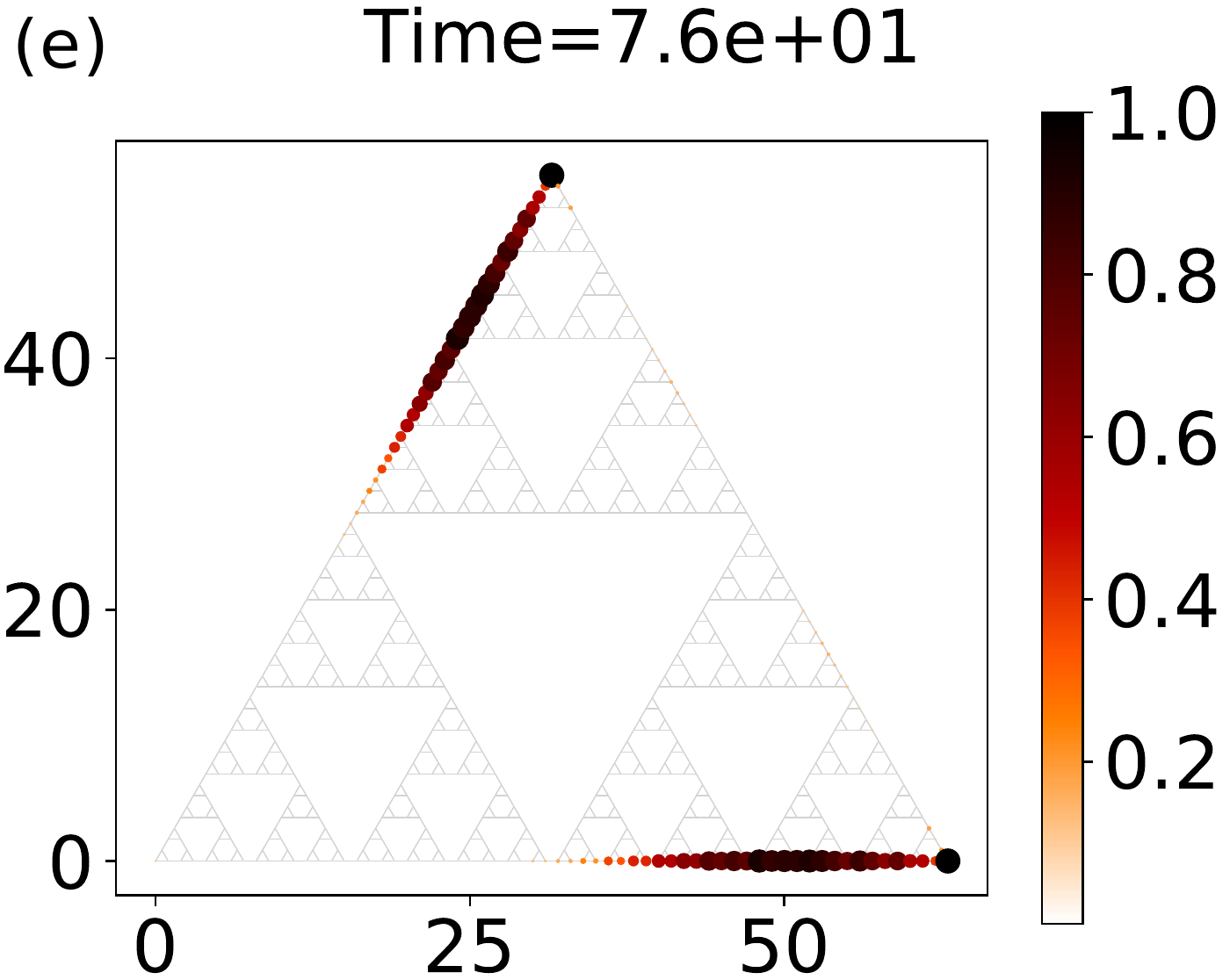}  
\includegraphics[scale=0.25]{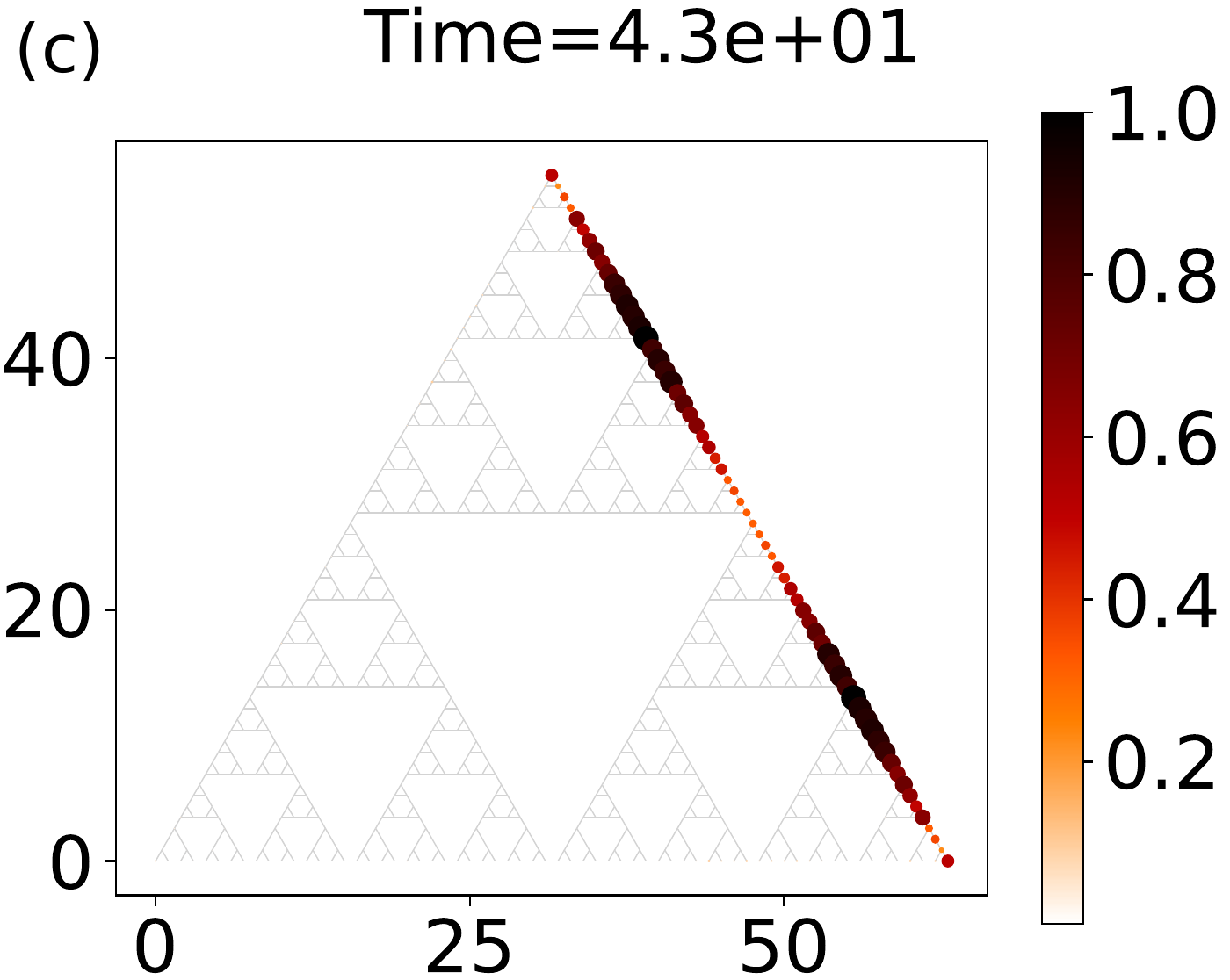}  
\includegraphics[scale=0.25]{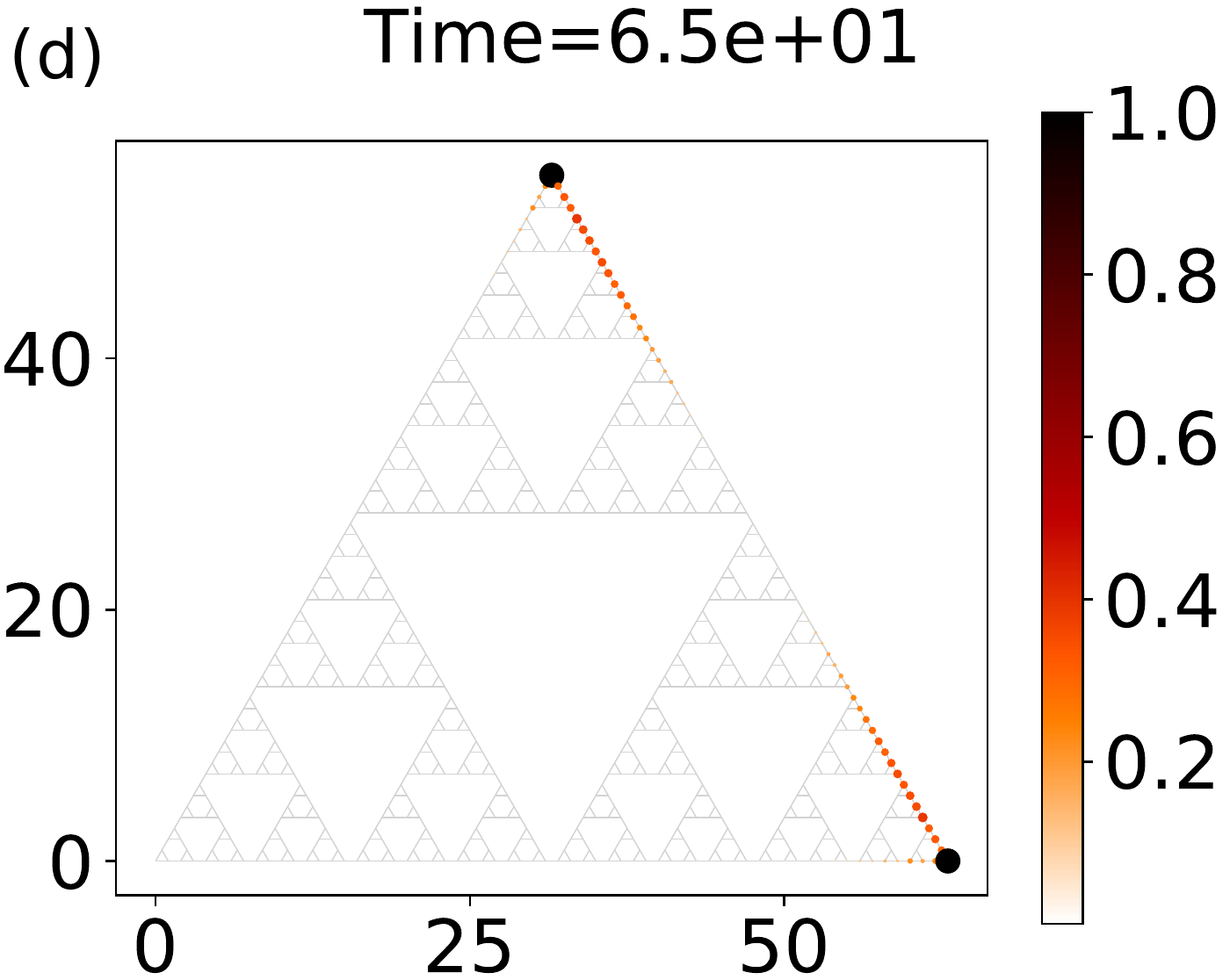}  
\caption{\footnotesize{Time evolution of a state (a$\rightarrow$b$\rightarrow$c$\rightarrow$d$\rightarrow$e$\rightarrow$f), initially localised in the $c$ orbital of one of the corner sites on SG-3 with $g=6$, evolved under $\hat{\mathbb{H}}~(t=0,~\lambda=1)$. The initial state is projected onto a sector defined by $-0.5<E<0$. The color bar represents the relative density per site of an eigenstate, $\ket{\psi_n}$, defined by $\rho_n(j) / {\max}(\rho_n(j))$.}}\label{QD_SG_3_t_0_M_0}
\end{figure}

\begin{figure}
\includegraphics[scale=0.25]{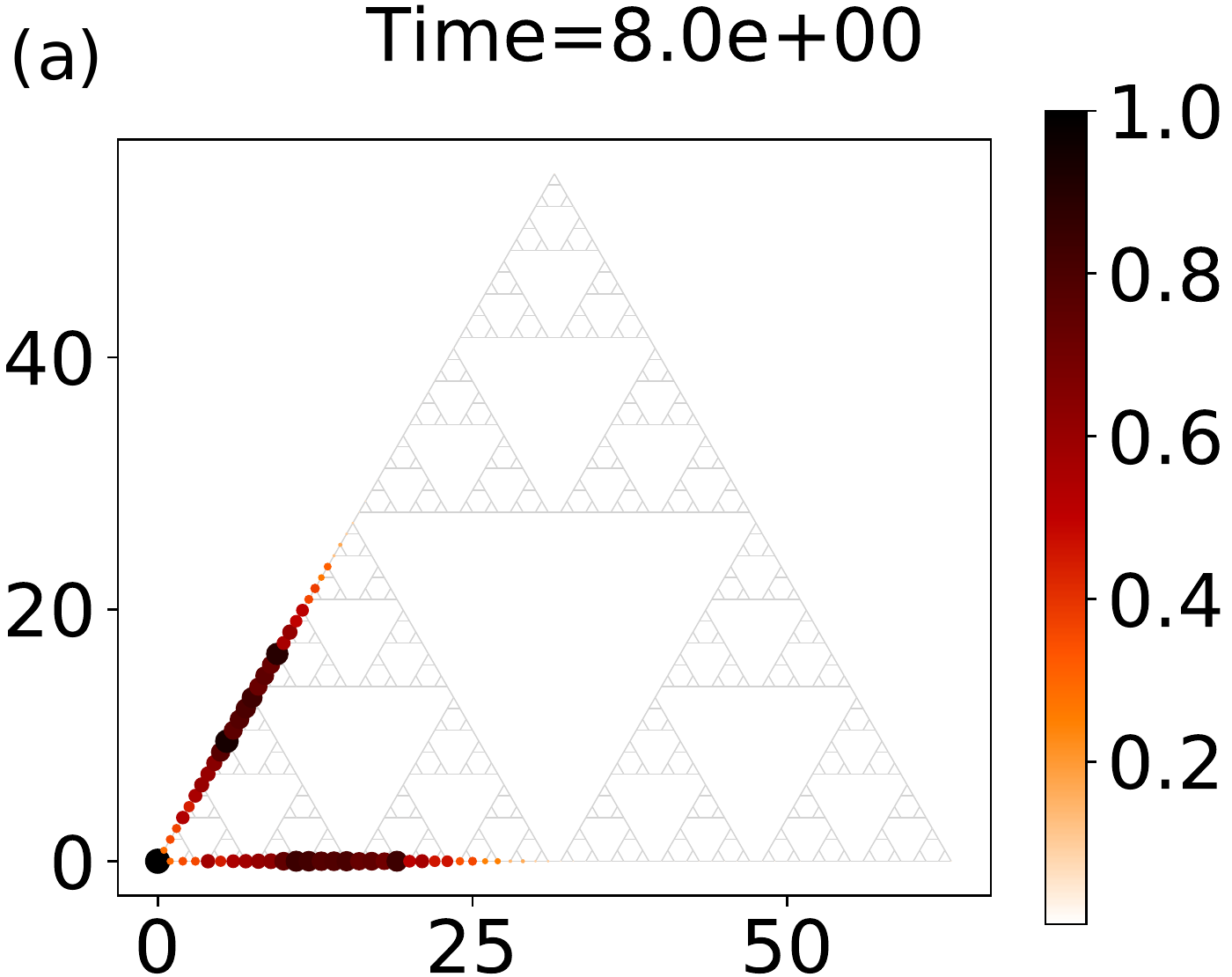}
\includegraphics[scale=0.25]{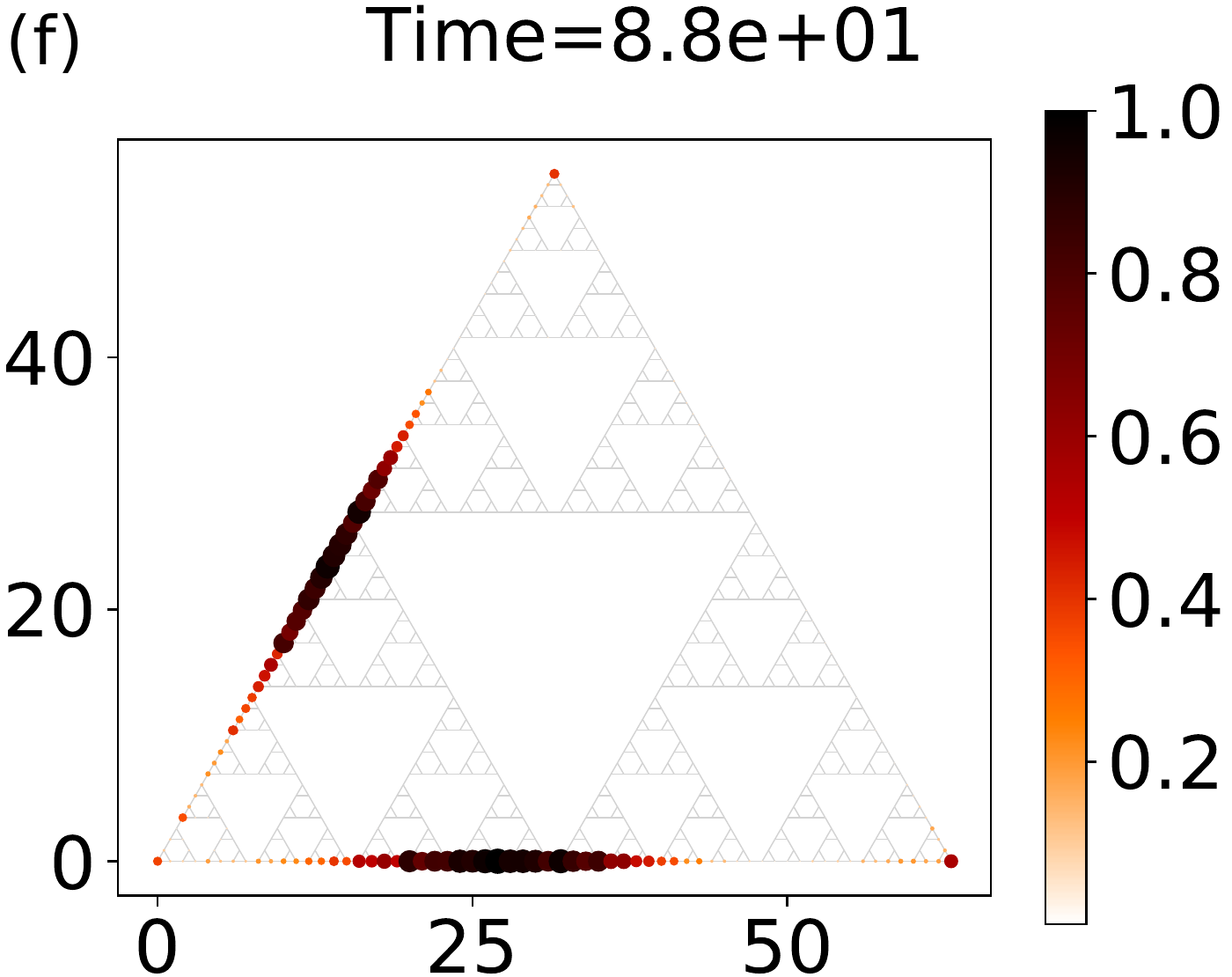}  
\includegraphics[scale=0.25]{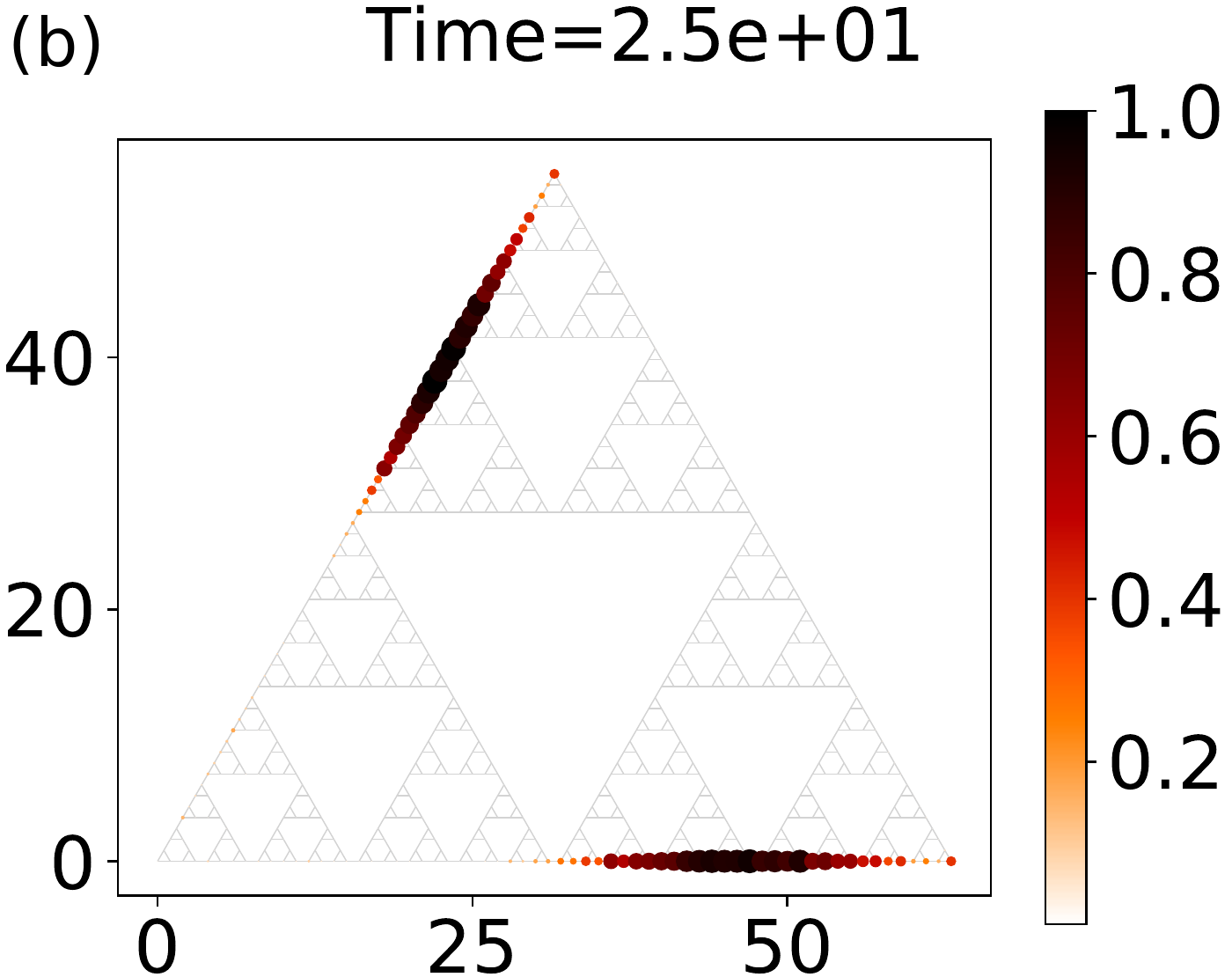}  
\includegraphics[scale=0.25]{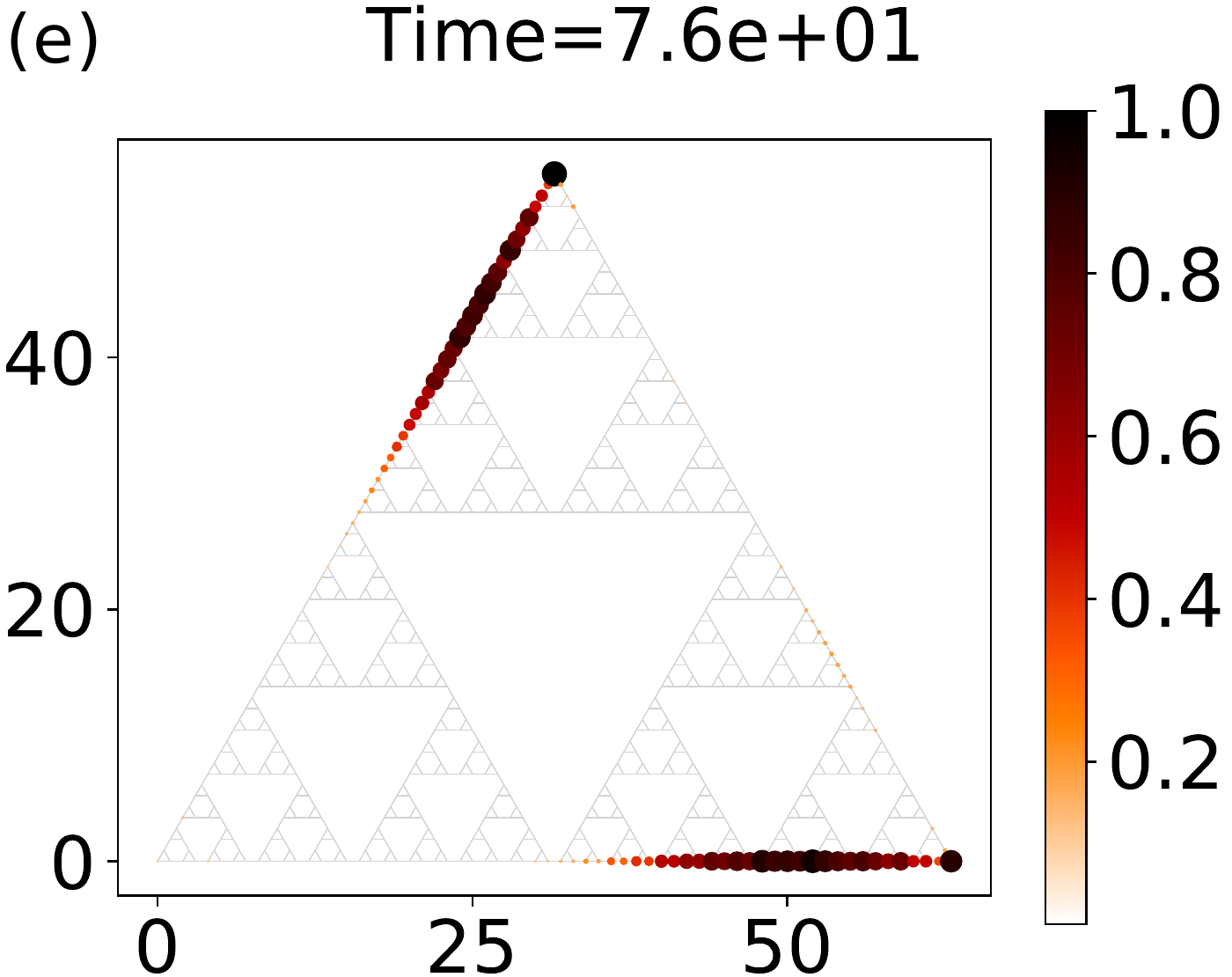}  
\includegraphics[scale=0.25]{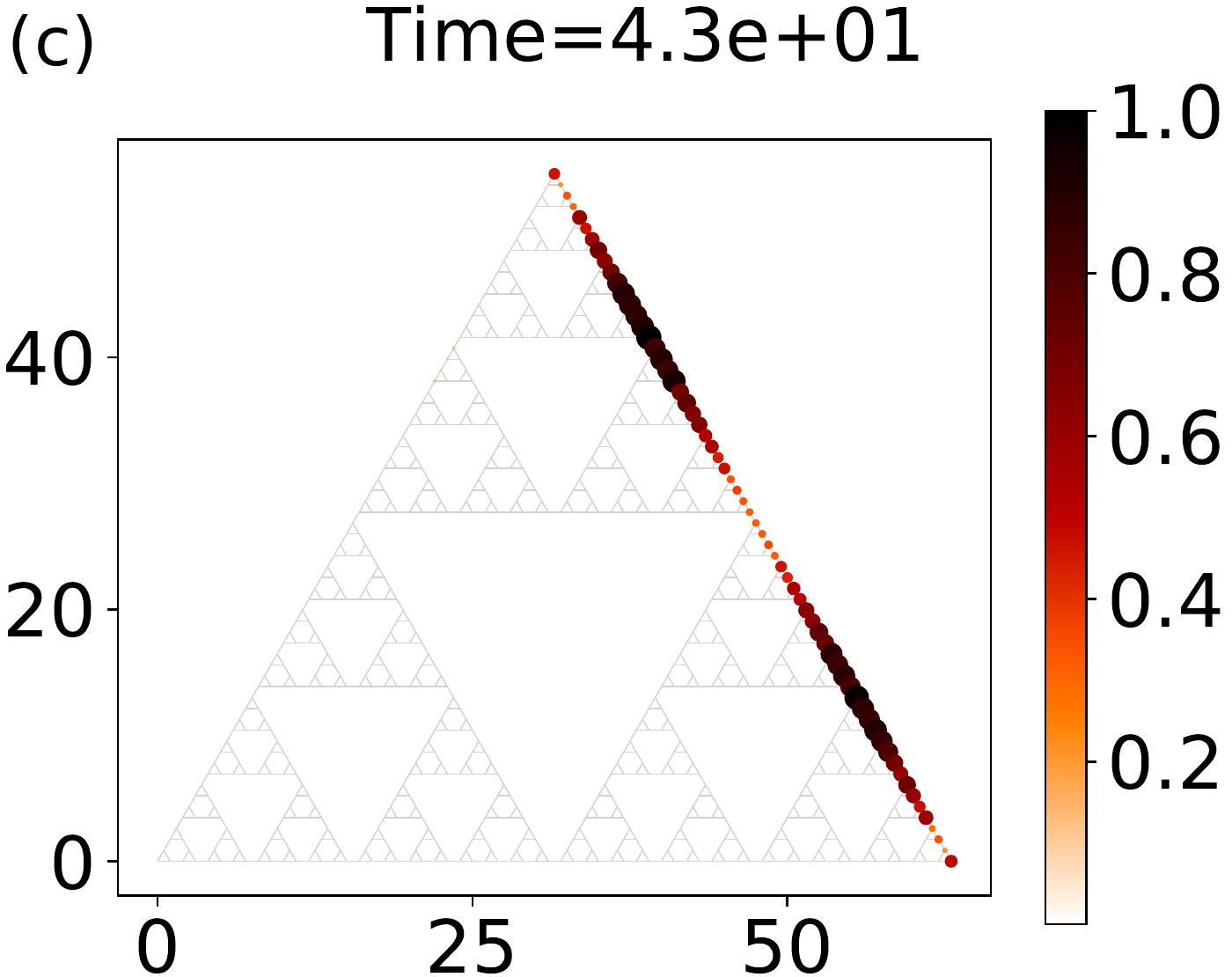}  
\includegraphics[scale=0.25]{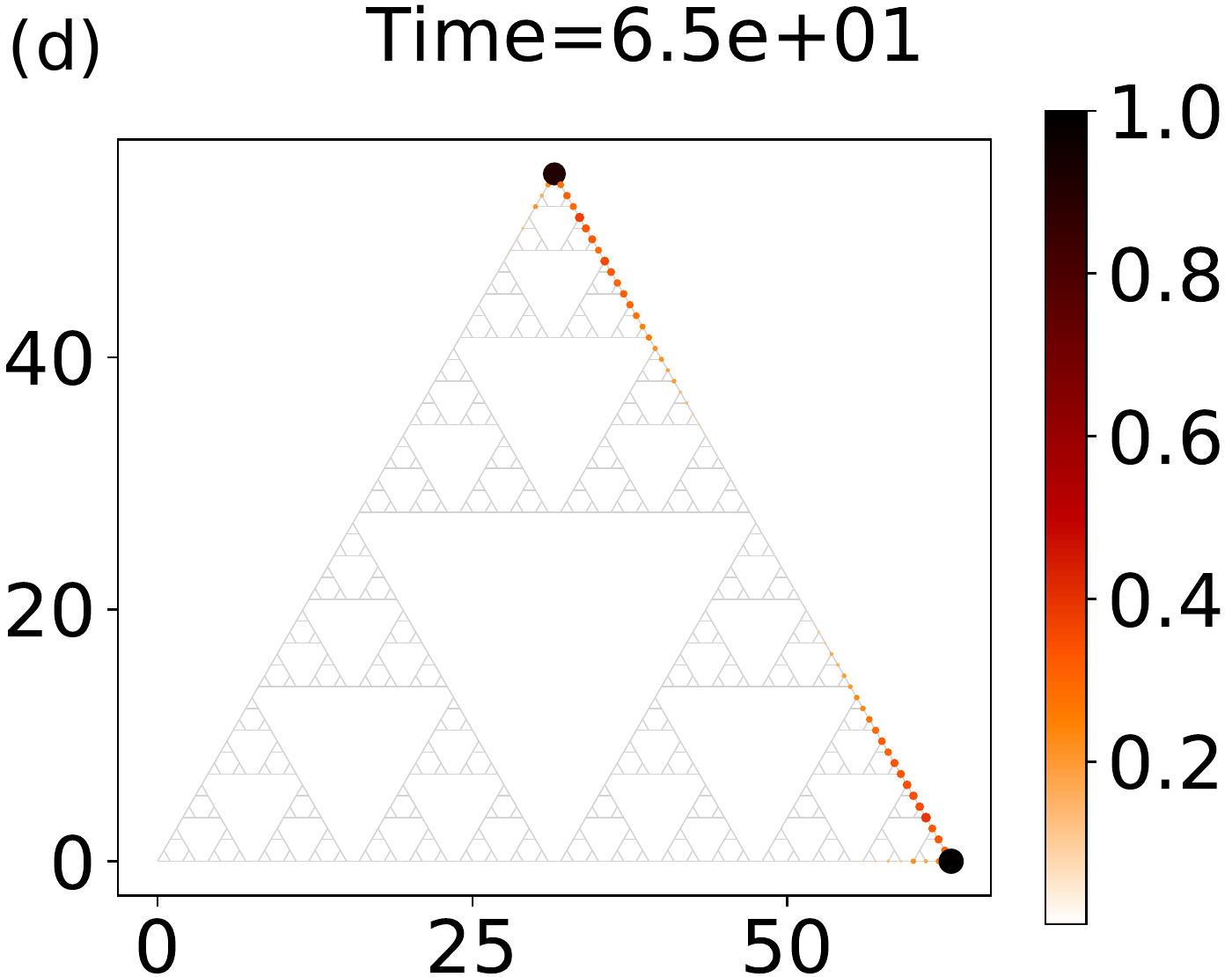}  
\caption{\footnotesize{Time evolution of a state
(a$\rightarrow$b$\rightarrow$c$\rightarrow$d$\rightarrow$e$\rightarrow$f), initially localised in the $c$ orbital of one of the corner sites on SG-3 with $g=6$, evolved under $\hat{\mathbb{H}}_{dis}~(t=0,~\lambda=1)$ with $W=0.1$. The initial state is projected onto a sector defined by $-0.5<E<0$. The color bar represents the relative density per site of an eigenstate, $\ket{\psi_n}$, defined by $\rho_n(j) / {\max}(\rho_n(j))$.}}\label{QD_W_0pt1_SG_3_t_0_M_0}
\end{figure}

We find that the spectrum of $H_{xy}$ on SG-4 hosts highly degenerate levels at the Fermi energy (Fig.\ \ref{Spectra_H_xy}(a)), which is not present in the case of SG-3. The Chern number for the collection of degenerate levels at $E_{F}$ turns out to be zero, when computed using Eq.\ \eqref{LCN}. On SG-3, $H_{xy}$ hosts doubly degenerate zero energy states. Interestingly, these zero energy states are edge-like states, completely localized on the sites present on the triangle of the $1$st generation. In fact, we observe that all states close to zero energy, shown in the inset of Fig.\ \ref{Spectra_H_xy}(b), are edge-like states. A few examples of such states are shown in Fig.\ \ref{Wv_SG_3_t_0_M_0}. In this case also, we find the Chern number to be zero, when computed by projecting onto the filled states (half-filling). However, looking at the dynamics of the edge-like states close to the Fermi energy, we find two modes of opposite chirality being present in the system (shown in Fig.\ \ref{QD_SG_3_t_0_M_0}). We also check the wave-packet dynamics in presence of weak Anderson disorder (shown in Fig.\ \ref{QD_W_0pt1_SG_3_t_0_M_0}) and find this characteristic in the dynamics being robust to weak disorders. The presence of robust edge-states is a signature of a topologically non-trivial phase. So, $H_{xy}$ is topologically non-trivial on SG-3 and the Chern number being zero is merely a consequence of the time-reversal symmetry in the system.
\newline

Here, we would like to point out that $H_{xy}$ has a gapless spectrum on the square lattice and the triangular lattice, with Dirac cones at the high symmetry points of their respective Brillouin zones. Their corresponding Bloch Hamiltonians are given by, $H^{sq}_{xy}(\va{k})=\sin(k_x)\sigma_x+\sin(k_y)\sigma_y$ for the square lattice, and $H^{tri}_{xy}(\va{k})=2(\sin(k_{x})+\sin(k_x/2)\cos(\sqrt{3}k_y/2))\sigma_x+2\sqrt{3}\cos(k_x/2)\sin(\sqrt{3}k_y/2)\sigma_y$ for the triangular lattice. As these systems are not gapped, these do not fall under the usual classification of gapped topological phases in terms of the tenfold symmetry classes. The system has time-reversal symmetry ($T^2=-1$) and hence has Kramer's degeneracy, thus preventing chiral dynamics in the system.
\newline

In 2-dimensional two band Chern insulators (absence of time-reversal symmetry), the forward and the backward moving modes are localized on edges which are spatially separated and this prevents the possibility of scattering between them. However, in the presence of Kramer's degeneracy, each edge-mode is accompanied by its Kramer's degenerate counterpart which moves in the opposite direction on the same edge. So the scattering between the Kramer's pairs cannot be prevented unless there is an additional spin (or spin-like) degree of freedom to couple to the edge-modes, thus making them helical. Considering $H_{xy}$ on 2-dimensional translationally invariant systems, $H_{xy}$ puts two orbitals on each lattice site, thus making it a two band model if the underlying motif is a Bravias lattice. Hence, $H_{xy}$ does not have any additional spin (or spin-like) degree of freedom and no chiral or helical edge dynamics can be observed for $H_{xy}$ on square or triangular lattices.
In this context, the wavepacket dynamics of $H_{xy}$ on SG-3 is particularly interesting. Here, the two counter propagating edge-like modes shown in Fig.\ \ref{QD_SG_3_t_0_M_0} do not scatter among themselves even in the presence of disorder (Fig.\ \ref{QD_W_0pt1_SG_3_t_0_M_0}). Notice that the arguments used earlier to describe the edge-state dynamics of $H_{xy}$ on square and triangular lattices are no longer valid for self-similar systems due to lack of an equivalent picture for the band structure in this case. The fact that $H_{xy}$ neither shows such dynamics on 2-dimensional lattices nor on the other self similar structure, SG-4, but only on SG-3, suggests that such dynamics is due to the interplay between the self-similarity and the local coordination of SG-3. 
\newline

\begin{figure}
\includegraphics[scale=0.25]{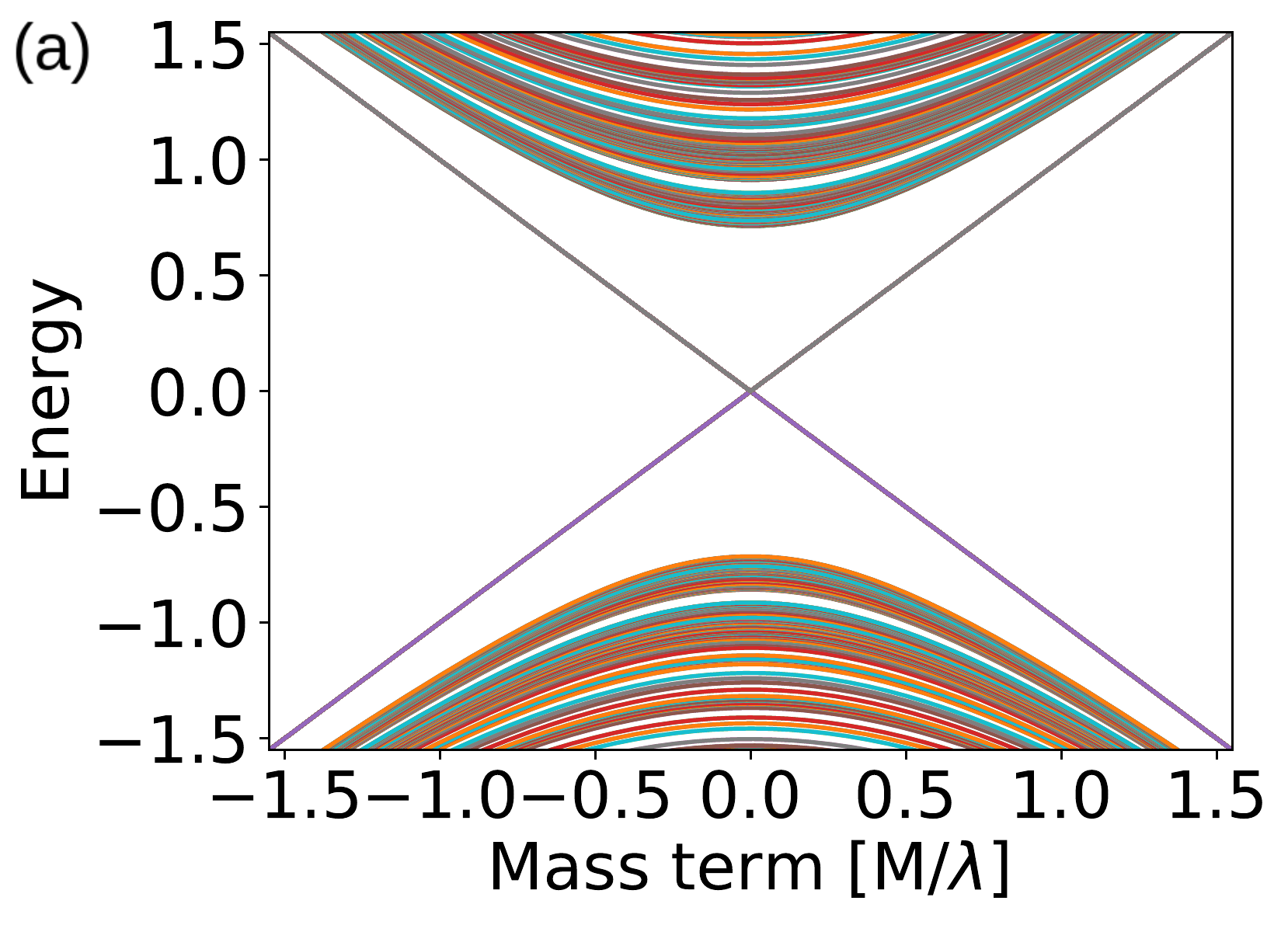}
\hfill
\includegraphics[scale=0.25]{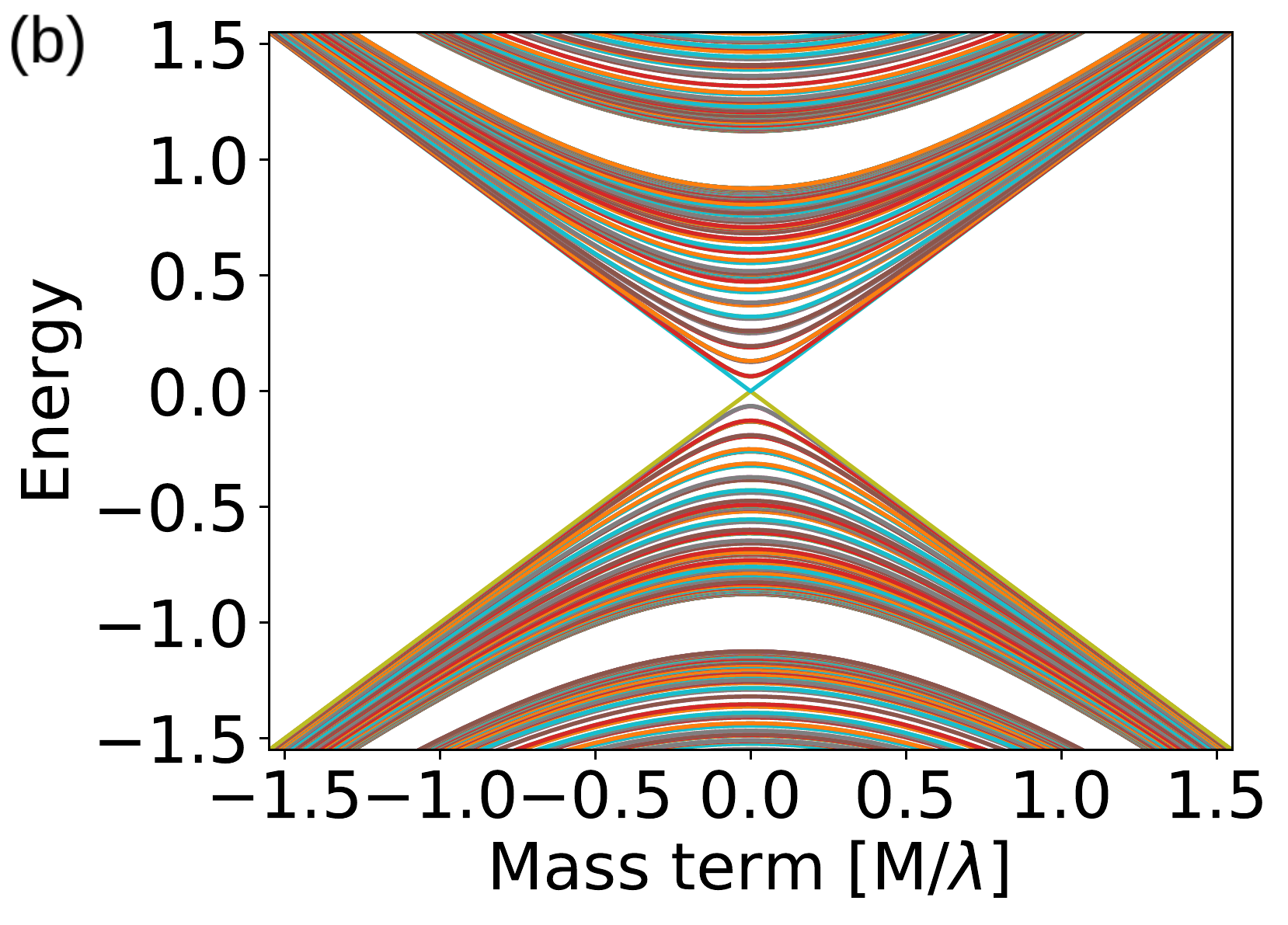}
  \caption{\footnotesize{Part of the spectra of $M\tau_{z}+H_{xy}$ as a function of $M$ on (a) SG-4 and (b) SG-3. The $M\tau_z$ term breaks the time-reversal symmetry and opens a gap proportional to $M$ in the spectrum of $H_{xy}$.}}\label{EvM_t_0}
\end{figure}

The $M\tau_z$ term creates a gap in the spectra of $\mathbb{H}$ (shown in Fig.\ \ref{EvM_t_0}) on both the structures.  For SG-4, the flatband at zero energy splits into two flatbands with energies $M$ and $-M$. Addition of a $M\tau_z$ term breaks the time-reversal symmetry of $\mathbb{H}$, since $T^{-1} (M\tau_z + H_{xy}) T = (-M\tau_z+ H_{xy})$. However, we still find the spectra of $\mathbb{H}$ on SG-3 to consist of doubly degenerate states as in the case of $H_{xy}$. This double-degeneracy is independent of the fractal structure and is due to non-spatial symmetries of $H_{xy}$ as shown in Appendix \ref{h_xy_degeneracy}.

\subsection{$t=\lambda \neq 0$}

\begin{figure}
\includegraphics[scale=0.18]{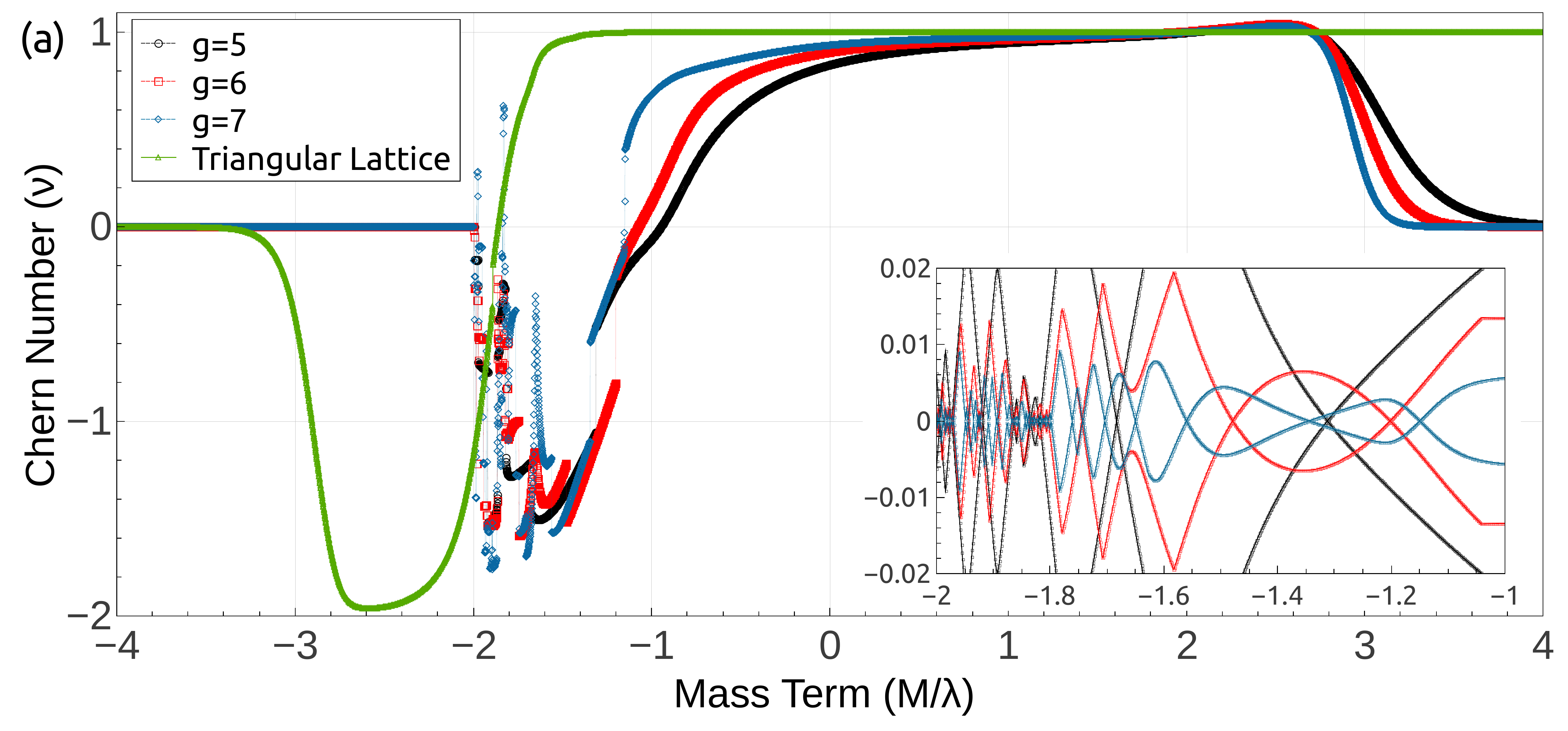}
\includegraphics[scale=0.18]{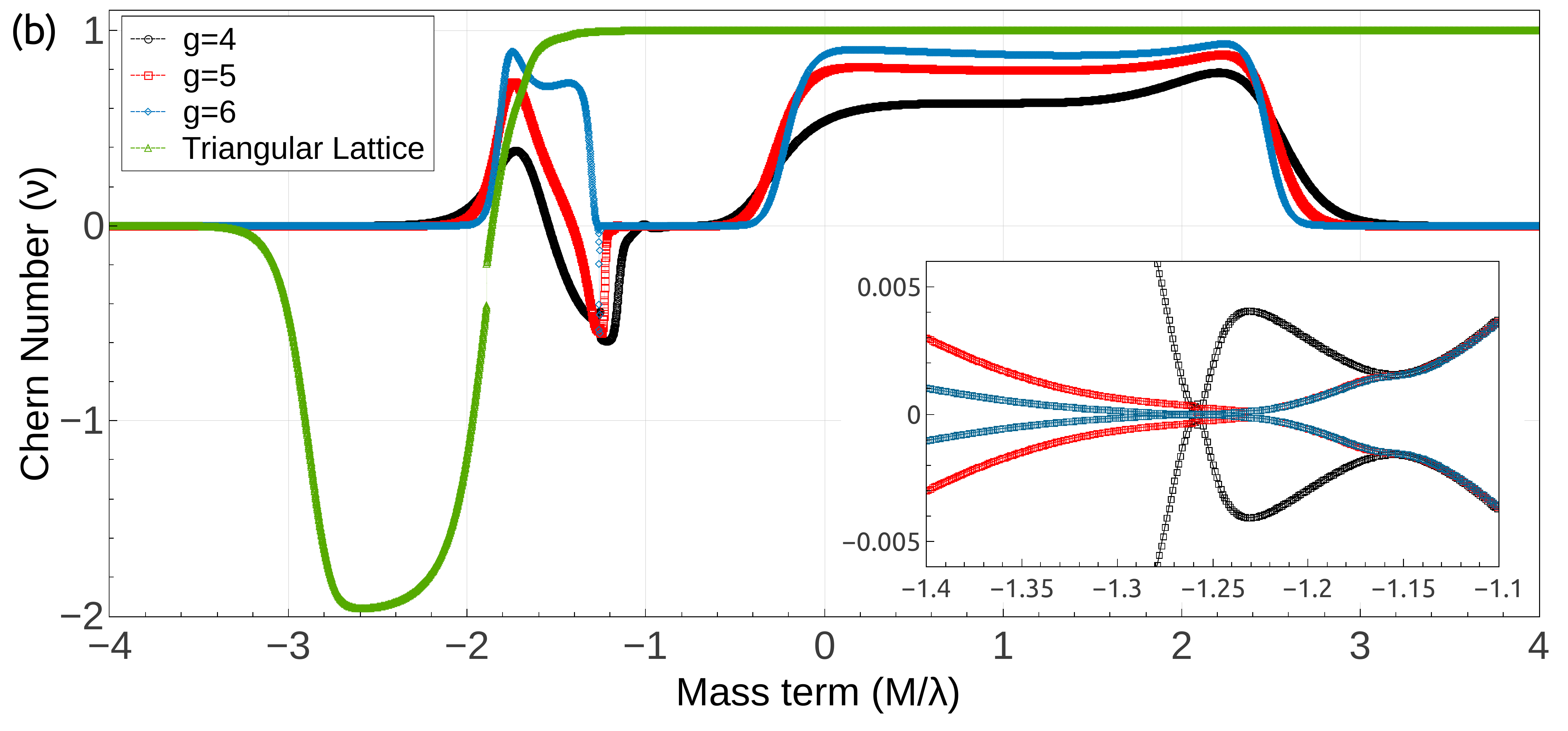}  
\caption{\footnotesize{Real space Chern number for $\mathbb{H}$ in the regime $\lambda=t$ on (a) SG-4 and (b) SG-3. The computation is done using Eq.\ \eqref{LCN}, which strongly depends on the system size. We do a system size scaling by looking at Chern numbers for different generations $g$. The inset in each plot shows the first two energy levels closest to the Fermi energy, for different generations. The legend for the insets are the same as that for the main plots. The inset of (a) shows numerous level crossings for SG-4, which increase with $g$. The inset of (b) shows a single level crossing for SG-3. }}\label{chern_l=t}
\end{figure}

Switching on both [$c\rightarrow c$, $d\rightarrow d$] and [$c\rightarrow d$, $d\rightarrow c$] hoppings brings a lot of interesting physics into the picture. From Fig.\ \ref{chern_l=t}, we find that $\mathbb{H}$ hosts topological phases on both the structures. In the regime $0\lesssim (M/\lambda) \lesssim 2.5$, both SG-3 and SG-4 host topological phases with the same Chern number $\nu=1$ and support edge-like states. For SG-4 with different boundary conditions, similar edge-like states were reported \cite{Agarwala2018}, which were robust against random onsite disorder, and possessed a chiral nature. In our case also, we find the same for both SG-3 and SG-4 in this regime.  
\newline

To serve as a reference for studying the real space Chern number computations, we also compute the Chern number for the model on a triangular lattice using Eq.\ \ref{LCN}, with a system size comparable to that of the fractals. The results are shown as the green curve in Fig.\ \ref{chern_l=t}(a) and (b). Due to the strong dependence of Eq.\ \ref{LCN} on the system size, the transitions from one Chern number to the other is not very sharp. So the real space Chern number is only strongly quantized away from the transition region. A detailed numerical computation on the strength of the quantization of the real space Chern number on crystal lattices is presented in Fig. 4 in reference \cite{Bianco2011}.
\newline

\begin{figure}
\includegraphics[scale=0.3]{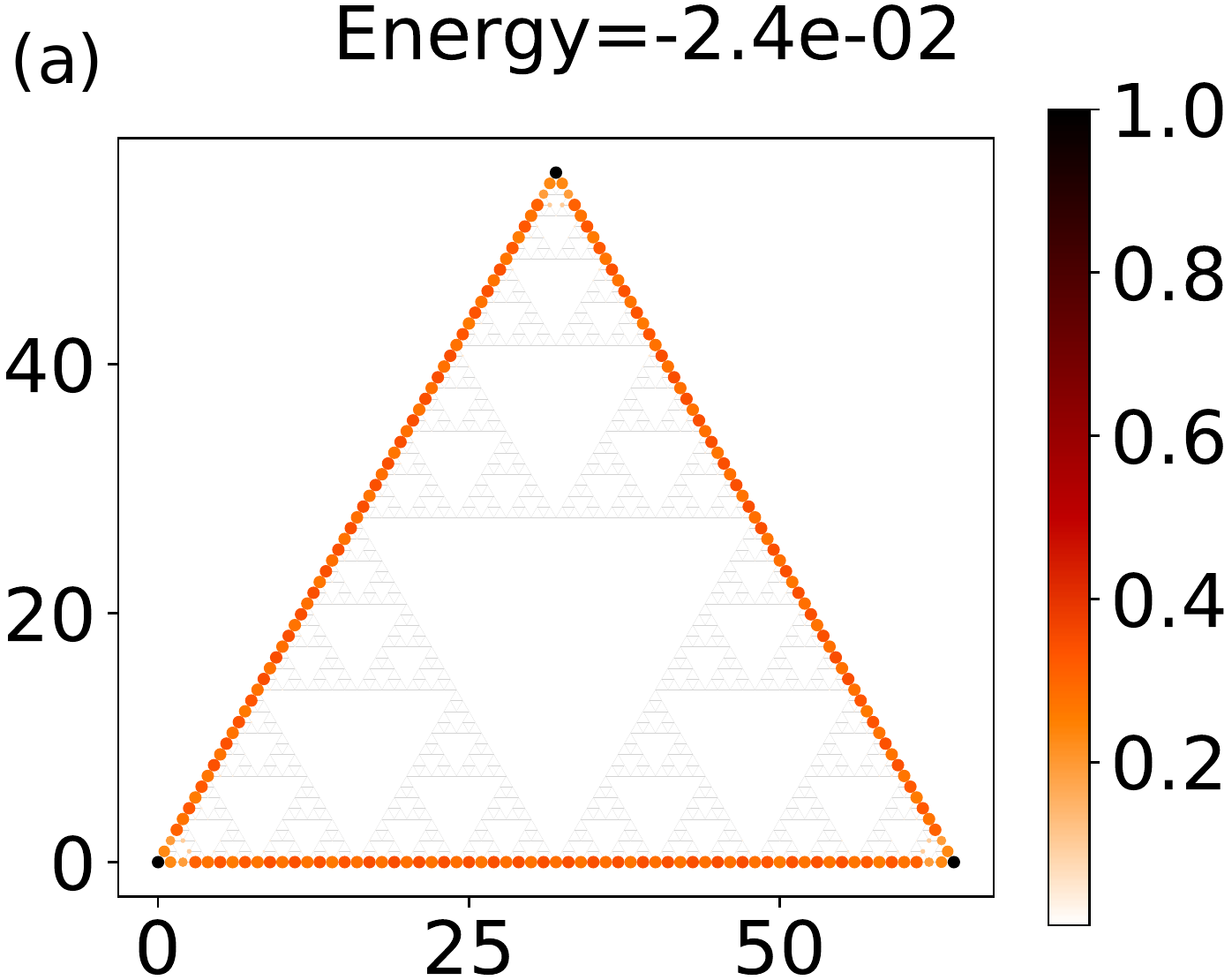}
\includegraphics[scale=0.3]{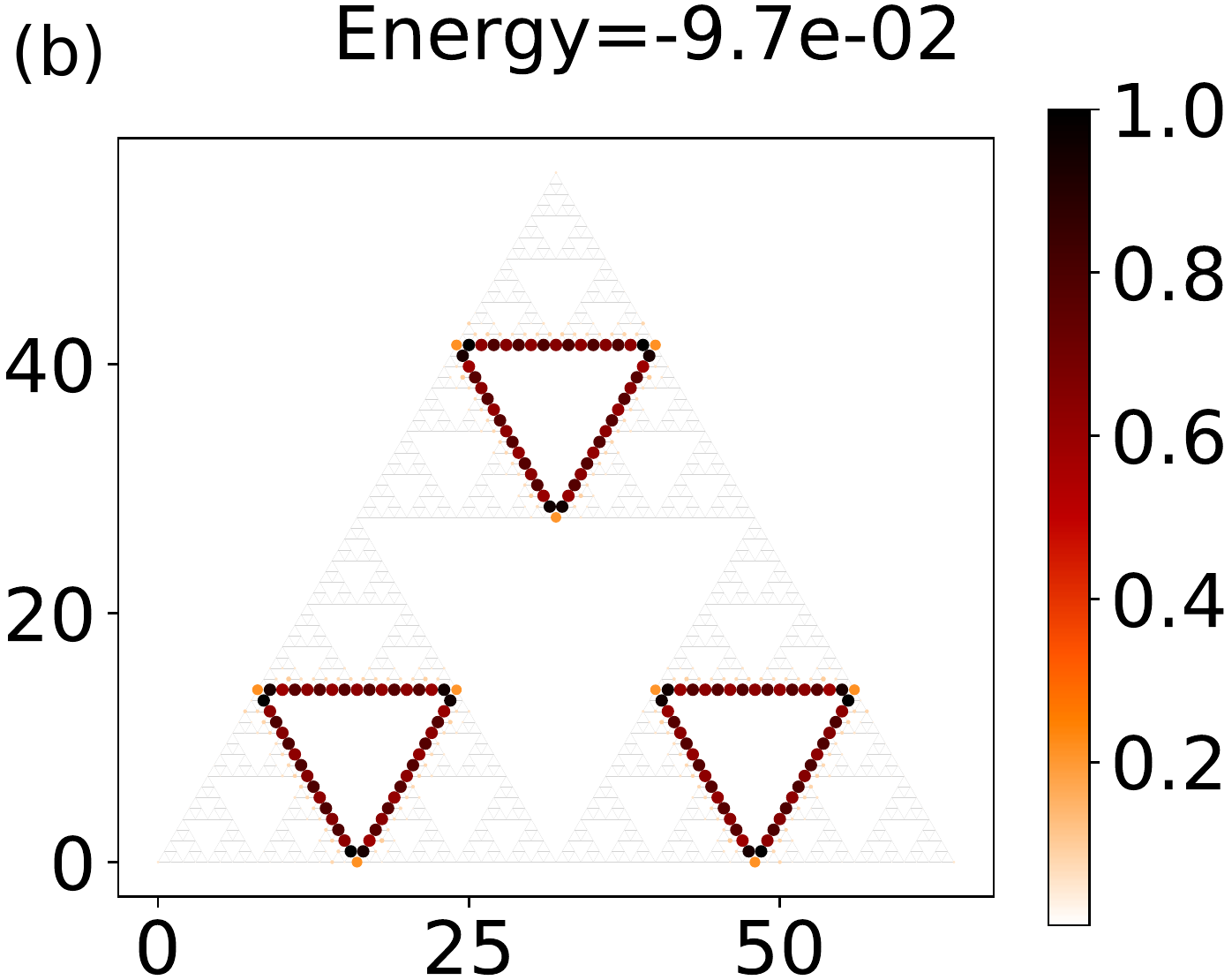}
\includegraphics[scale=0.3]{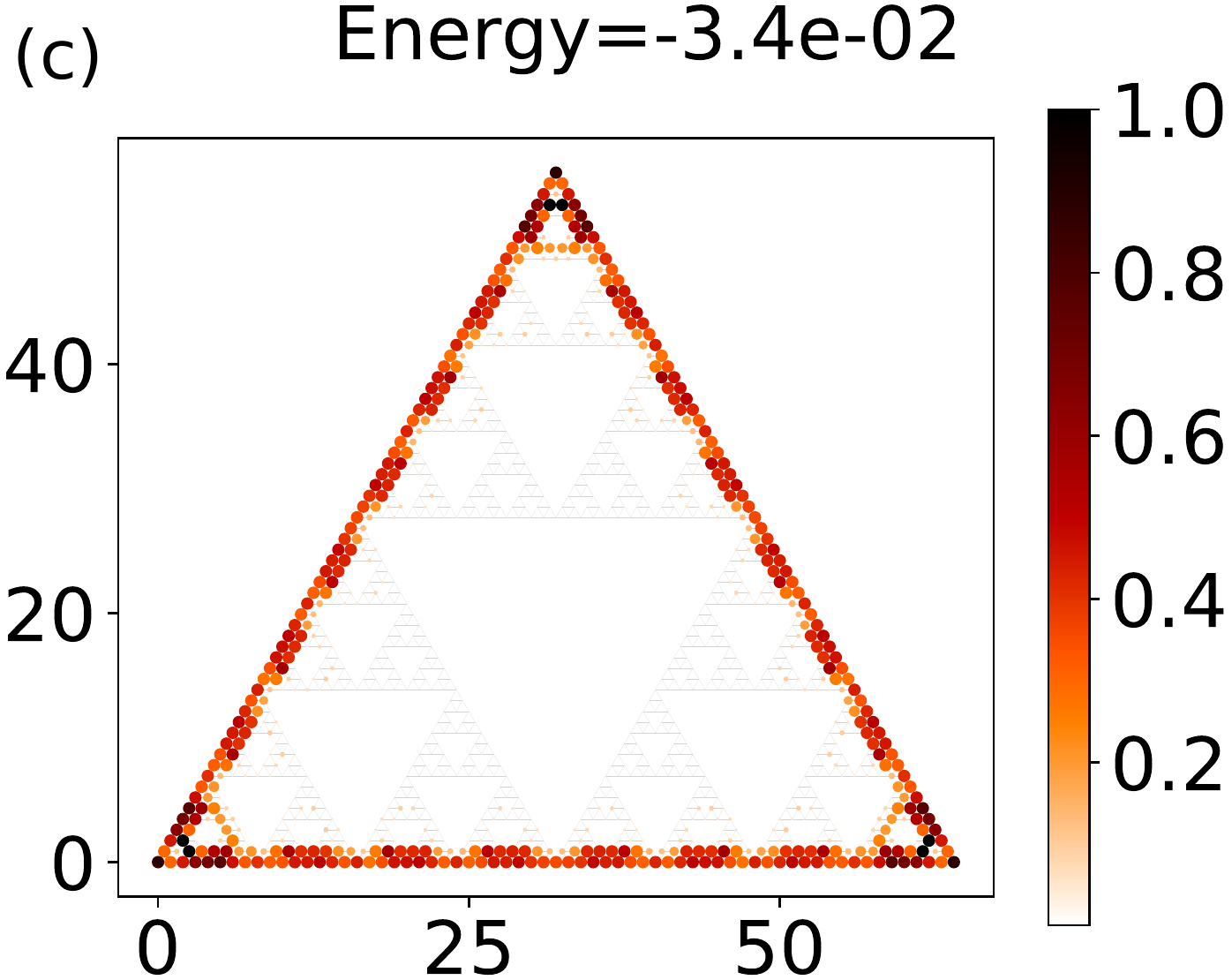}
\includegraphics[scale=0.3]{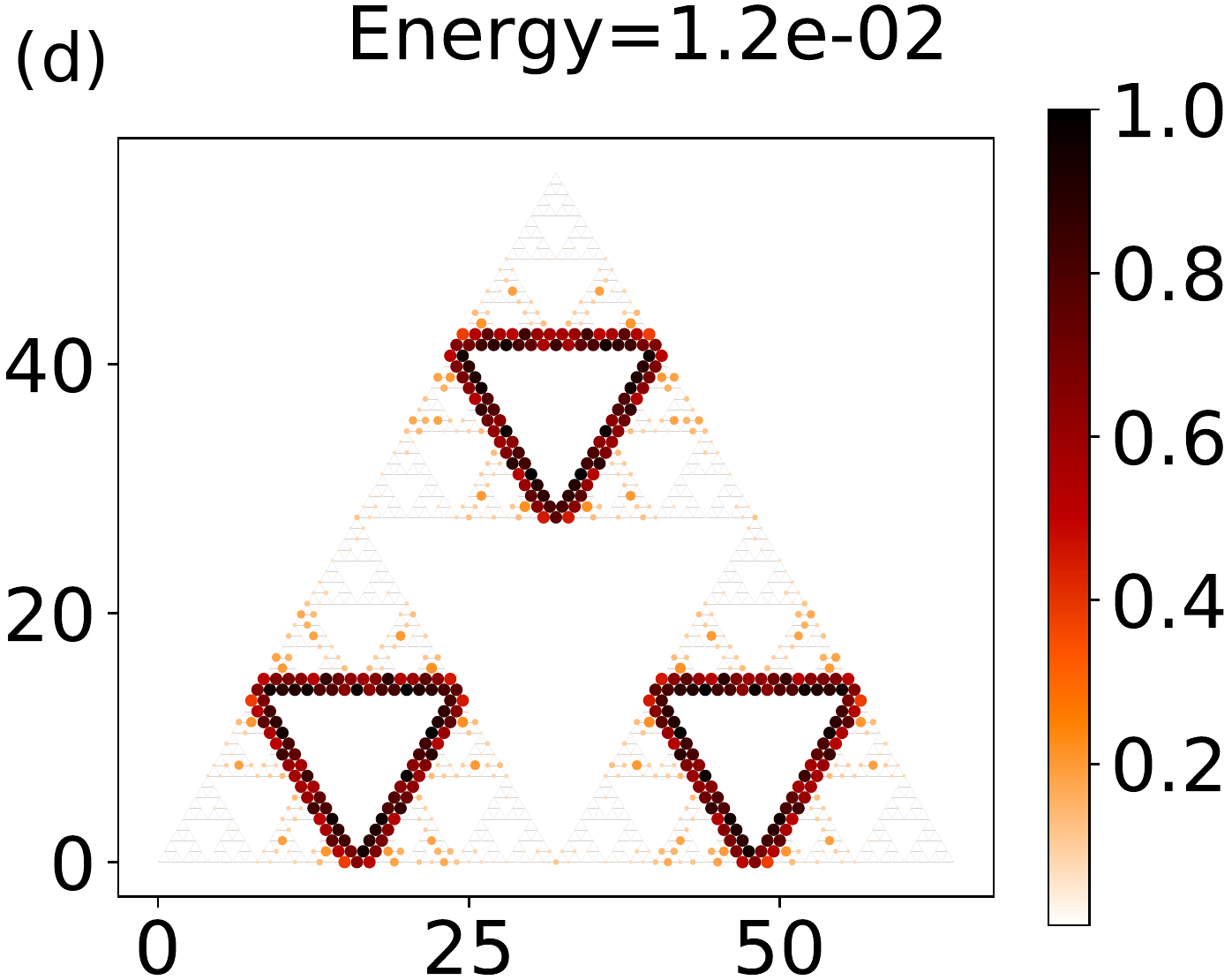}
\caption{\footnotesize{Edge-like states on SG-4. (a) and (b) are examples of edge-like states for $-1.6 \lesssim (M/\lambda) \lesssim -1.3$ and, (c) and (d) are examples of edge-like states for $0 \lesssim (M/\lambda) \lesssim 2.5$. Notice the difference in the localization pattern of edge-like states in the two regimes. For $0 \lesssim (M/\lambda) \lesssim 2.5$, the states are localized on a single layer of sites which enclose the triangles of a particular generation. In contrast, for $-1.6 \lesssim (M/\lambda) \lesssim -1.3$, the states are primarily localized on two consecutive layers of sites which enclose the triangles of a particular generation.   
}}\label{edge_states_SG-4}
\end{figure}

\begin{figure}
\includegraphics[scale=0.19]{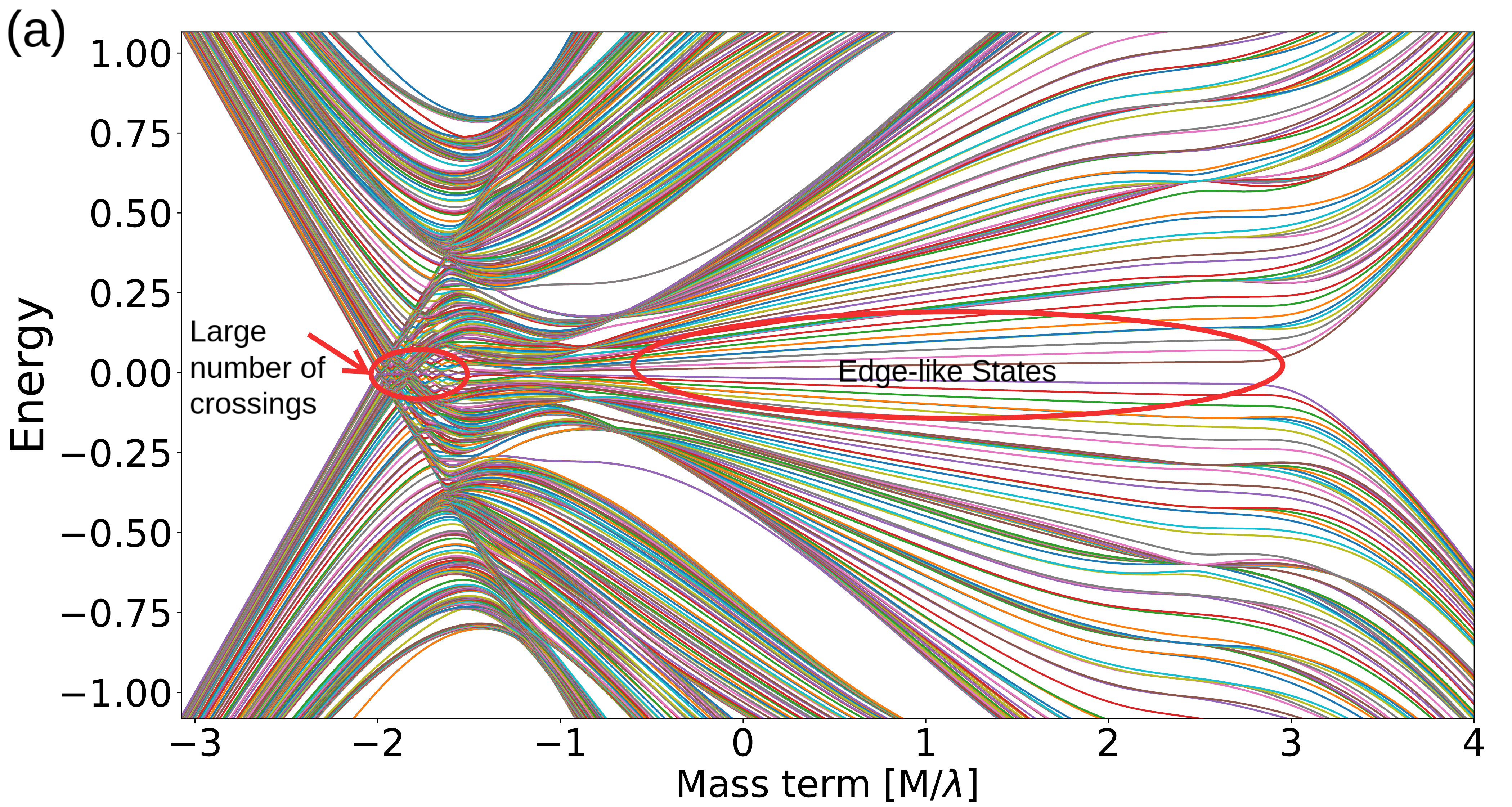}
\includegraphics[scale=0.30]{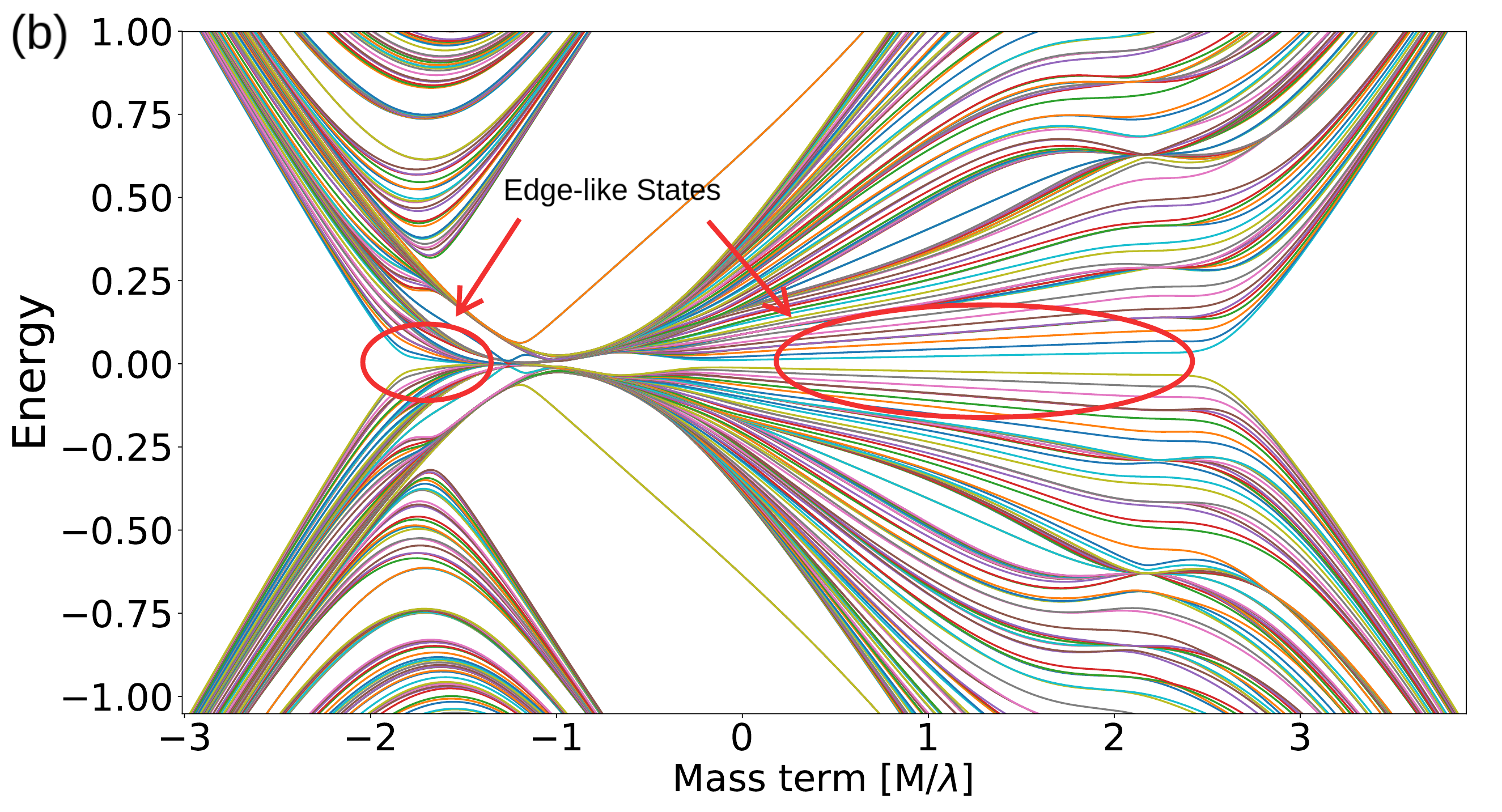}  
\caption{\footnotesize{Part of the spectra for $\mathbb{H}$ in the regime $\lambda=t$ on (a) SG-4 with $g=7$, and (b) SG-3 with $g=6$. The regions of the spectra which host edge-like states are pointed out for both the structures. These regions correspond to the topological regions in Fig\ \ref{chern_l=t}. }}\label{spec_l=t}
\end{figure}

In the regime, $-2 \lesssim (M/\lambda) \lesssim -1.2$, SG-3 and SG-4 host different topological phases, characterized by different Chern numbers. For SG-4, in the regime, $-1.6 \lesssim (M/\lambda) \lesssim -1.3$, where there are no level crossings, we find the Chern number to be transitioning towards $\nu=-2$. Although we do not see a good enough quantization of the Chern number numerically, we do find edge-like states and chiral wave-packet dynamics in this regime, suggesting that the phase is not trivial. Also, the localization pattern of edge-like states in this regime is different from that of the regime with $\nu=1$ (see Fig.\ \ref{edge_states_SG-4}), suggesting $\nu=-2$ as opposed to $\nu=-1$ for this regime. For SG-4, there are many level crossings in the regime $-2\lesssim (M/\lambda) \lesssim -1.6$ (Fig.\ \ref{spec_l=t}). The number of level crossings increases with generation of the fractal. Given that the Chern number is not well defined at level crossings, the computation using Eq.\ \eqref{LCN} does not give a definitive value (Fig.\ \ref{chern_l=t}(a)). 
\newline

For SG-3, there is exactly one level crossing at $(M/\lambda) \approx -1.24$, which seems to be one of the transition points from a topological phase to a trivial phase. In the regime, $-2 \lesssim (M/\lambda) \lesssim -1.24$, we find a topologically non-trivial phase with $\nu=1$ on SG-3, which is different from what we found for SG-4. Although the Chern number computation for smaller generations shows a small dip around $(M/\lambda) \approx -1.24$ (see the red and black curve in Fig.\ \ref{chern_l=t}(b)), this dip vanishes as we do the computation for higher generations (blue curve in Fig.\ \ref{chern_l=t}(b)). This shows that for $t=\lambda \neq 0$, SG-3 hosts a trivial phase and only one  topological phase with $\nu=1$ in the thermodynamic limit.

\section{Conclusion}\label{Conclusion}

We have explored the properties of a geometry dependent Hamiltonian on two different finite fractal structures (SG-3 and SG-4) which only differ in the way the sites are coordinated. The Hamiltonian has different non-spatial symmetries for different parameter regimes. We study the systems in each of these parameter regimes separately. We find that the topological properties of this Hamiltonian are significantly different on the two structures.
\newline

In the regime $t=\lambda\neq0$, where only charge-conjugation symmetry is present, the \textit{half-BHZ} model can host both topologically trivial and non-trivial phases characterized by a non-zero real-space Chern number, on both the structures. For both SG-3 and SG-4, we find chiral edge-like eigenstates close to the Fermi energy for the parameter regimes corresponding to the topologically non-trivial phases. However, the phases obtained for each of the structures are different which is evident from their respective plots of Chern number (Fig.\ \ref{chern_l=t}). In the regime $\lambda\neq{t=0}$, where all the three symmetries (time-reversal, charge-conjugation, and orbital symmetry) are present, we find the existence of non-trivial doubly degenerate edge-like eigenstates with opposite chiralities near the Fermi energy on  SG-3. No such chiral edge-like states are present in case of SG-4 for this particular parameter regime. Instead, a highly degenerate zero energy band is present in SG-4 which we expect to be topologically trivial. The existence of doubly degenerate robust edge-like states of opposite chirality on SG-3 is particularly interesting. This leads to unexpected wavepacket dynamics in which two counter propagating edge-like modes do not scatter into each other. Such dynamics is not present when the model studied on the square and triangular lattices, and SG-4. This highlights the role of coordination in determining the physics on self-similar structures.   
\newline
 
As the distinguishing factor between the two structures is their coordination number, we arrive at the conclusion that the topological properties on self-similar lattice systems depend significantly on the way the sites are coordinated. The description of topological phases in translationally invariant non-interacting systems does not explicitly take the coordination into account. There, coordination is taken implicitly into account in the matrix elements of the corresponding Bloch Hamiltonians. But that is not possible for systems which lack translational symmetry. The results of this work suggest that, in order to extend the present classification scheme, it is important to use a framework which explicitly takes the coordination of sites into account. Perhaps one way to look at such systems is to use a framework of graphs.

\begin{acknowledgments}
We thank Aniket Patra, Adhip Agarwala and Blazej Jaworowski for useful discussions.
\end{acknowledgments}

\appendix

\section{Two-fold Degeneracy in $M\tau_z + H_{xy}$}\label{h_xy_degeneracy}
The two-fold degeneracy in $H_{xy}$ is a consequence of the fact that $T^{-1}H_{xy}T=H_{xy}$. Adding a mass term, $M\tau_z$, breaks this symmetry. However, eigenstates of $M\tau_z+H_{xy}$ still form degenerate pairs. Consider an eigenstate $\ket\psi$ of $H_{xy}$ with eigenvalue $\epsilon$. Due to the symmetry $\tau_z H_{xy}\tau_z=-H_{xy}$, we have that $\tau_z\ket\psi$ is also an eigenstate of $H_{xy}$ but with eigenvalue $-\epsilon$. Notice that addition of the $M\tau_z$ term also breaks this symmetry. Here, we analytically show that the effect of the $M\tau_z$ term is to hybridize $\ket\psi$ and $\tau_z\ket\psi$.
\newline

We assume an ansatz eigenstate of $M\tau_z+H_{xy}$ of the form $\alpha\ket\psi + \beta\tau_z\ket\psi$, with eigenvalue $E$. We have the following equation
\begin{multline}\label{degen_hybrid}
\Big( M\tau_z+H_{xy}\Big)(\alpha\ket\psi + \beta\tau_z\ket\psi)=\\ \Big(\beta M + \alpha\epsilon\Big) \ket\psi + \Big(\alpha M - \beta\epsilon\Big) \tau_z\ket\psi= E (\alpha\ket\psi + \beta\tau_z\ket\psi).
\end{multline}
For $\epsilon\neq 0$, we have that $\ket\psi$ and $\tau_z\ket\psi$ are orthogonal because they are eigenstates of $H_{xy}$ with different eigenvalues. Defining $\epsilon '= \epsilon /M$ and $E'=E/M$, and equating the coefficients of $\ket\psi$ and $\tau_z\ket\psi$ in Eq.\ \eqref{degen_hybrid}, we get
\begin{equation}\label{eigen_val_comp}
\beta + \alpha \epsilon'=\alpha E' ~~~;~~~
\alpha- \beta \epsilon'=\beta E'.
\end{equation}
Solving the pair of equations in \eqref{eigen_val_comp} for $\alpha$, $\beta$ and $E$, we get
\begin{equation}\label{val_a_b}
\dfrac{\alpha_{\pm}}{\beta_{\pm}}= \epsilon' \pm\sqrt{1+\epsilon'^{2}}  ~~~;~~~
E_{\pm}=\pm\sqrt{M^2+\epsilon^{2}}.
\end{equation}
So we have shown that $\alpha_{\pm}\ket\psi + \beta_{\pm}\tau_z\ket\psi$ are eigenstates of $M\tau_z+H_{xy}$ with $\alpha_{\pm}, \beta_{\pm}$ satisfying Eq.\ \eqref{val_a_b}. 
\newline

Now, as $P^{-1}(M\tau_z+H_{xy})P=-(M\tau_z+H_{xy})$, with $P=\tau_x\mathcal{K}$, we have that $P\ket\Phi$ is an eigenstate of the Hamiltonian, $M\tau_z+H_{xy}$, with eigenvalue $-\xi$, if $\ket\Phi$ is an eigenstate with eigenvalue  $\xi$. Hence, $\ket{\Psi_{+}}=\alpha_{+}\ket\psi + \beta_{+}\tau_z\ket\psi$ and $P\ket{\Psi_{-}}=P\Big(\alpha_{-}\ket\psi + \beta_{-}\tau_z\ket\psi\Big)$ are both eigenstates of the Hamiltonian, $M\tau_z+H_{xy}$, with the same eigenvalue $E_{+}$. Notice that the states, $\ket{\Psi_{+}}$ and $P\ket{\Psi_{-}}$, are orthogonal to each other, as $\ket\psi$, $P\ket\psi$, $\tau_{z}\ket\psi$ and $P\tau_{z}\ket\psi=-T\ket\psi$, are mutually orthogonal. The states $\ket\psi$ and $P\tau_z\ket\psi$ are orthogonal to the states $P\ket\psi$ and $\tau_z \ket\psi$ as they are eigenstates of $H_{xy}$, a Hermitian operator, with different eigenvalues. $P\ket\psi=-T\tau_z \ket\psi$ and $\tau_z \ket\psi$ are orthogonal as $\ket\psi$  and $T\ket\psi$ are orthogonal to each other as shown in section \ref{Tsym}. Hence $\ket{\Psi_{+}}$ and $P\ket{\Psi_{-}}$ cannot be the same state. This shows the existence of degenerate pairs even after the addition of the symmetry breaking term, $M\tau_z$.  However, for $\epsilon=0$, which corresponds to the zero energy modes of $H_{xy}$, the above argument no longer holds as $\ket\psi$ and $\tau_z\ket\psi$ no longer need to be orthogonal. Instead $\ket\psi$ and $\tau_z\ket\psi$ are eigenstates with energies $\pm M$.

\end{document}